\documentclass[twocolumn]{aastex63}

\usepackage{amssymb}	
\usepackage{amsmath}	
\usepackage{array}
\usepackage{bm}
\usepackage{CJK}
\usepackage{graphicx}	
\usepackage{hyperref}
\usepackage{xifthen}
\usepackage{xspace}
\usepackage{natbib}

\newcommand{\aizw}[1]{\textcolor{black}{#1}}

\begin{document}
\begin{CJK*}{UTF8}{gbsn}
\title{Discovery of Jet-Bubble-Disk Interaction:\\ Jet Feedback on a Protoplanetary Disk via an Expanding Bubble in WSB 52 }

\author[0000-0001-8877-4497]{Masataka Aizawa}
\affiliation{College of Science, Ibaraki University, 2-1-1 Bunkyo, Mito, 310-8512, Ibaraki, Japan }

\author[0000-0003-4039-8933]{Ryuta Orihara}
\affiliation{College of Science, Ibaraki University, 2-1-1 Bunkyo, Mito, 310-8512, Ibaraki, Japan }
\affiliation{Department of Astronomy, Graduate School of Science, The University of Tokyo, 7-3-1 Hongo, Bunkyo-ku, Tokyo 113-0033, Japan}

\author[0000-0002-3001-0897]{Munetake Momose}
\affiliation{College of Science, Ibaraki University, 2-1-1 Bunkyo, Mito, 310-8512, Ibaraki, Japan }

\begin{abstract}
While stellar jets and outflows are fueled by accretion from disks, their direct influence on disks remain unexplored. Here we revisit ALMA observations of \aizw{$^{12}\mathrm{CO}\,(J=2\mbox{--}1)$} line emission for the young stellar object WSB 52. We identify an expanding bubble that interacts with its protoplanetary disk. Given that the disk axis points toward the bubble center and the kinetic energy of the bubble is roughly $10^{41}$~erg, we postulate that stellar jets, aligned with the disk axis, have triggered the bubble. The bubble morphology is consistent with uniform expansion with partial concavity, implying the bubble-disk interaction. Correspondingly, the shape and the velocity field of protoplanetary disk appear to be deformed and exhibit high-velocity components, suggesting strong interactions and mass loss from the disk. The discovery of jet feedback onto the disk via the bubble---which we term the jet-bubble-disk interaction---sheds new light on the dynamical processes governing star and planet formation. 
\end{abstract}

\section{Introduction}
Protoplanetary disks are circumstellar disks composed of gas and dust, serving as the birthplaces of planets  \citep[e.g.,][]{hayashi1985,williams2011}. Recent high-resolution observations by the Atacama Large Millimeter/submillimeter Array (ALMA) have revealed the detailed distributions of disk materials, greatly enhancing our understanding of planet formation \citep[e.g.,][]{brogan2015,andrews2018, oberg2021}. The discovery of numerous substructures—such as gaps and rings—demonstrates that ongoing physical processes within the disk, including planetary formation, are actively taking place \citep[e.g.,][]{bae2022}. Moreover, comparisons of Class I and Class II disks, where fewer annular substructures have been identified in the earlier phases, offer insights into how protoplanetary disks evolve \citep{andrews2018,ohashi2023}. 

The materials of protoplanetary disks are supplied by larger-scale envelopes or cores and, in turn, promote mass accretion onto protostars.  A fraction of the inflowing material does not end up on the star but is instead expelled as jets and outflows, carrying away excess mass and angular momentum \citep[e.g.,][]{Blandford1982,matzner2000,machida2014}. Indeed, there is a strong correlation between the mass-loss rates and the protostellar accretion rates \citep[e.g.,][]{Hartigan1995}  \aizw{(see also Fig. 14 in \cite{Ellerbroek2013}) or \cite{lee2020} for review.)} The kinetic energy of ejected materials can be released into surrounding environments, increasing turbulence strengths \aizw{\citep[e.g.,][]{Norman1980,li2006,Nakamura2007}} and thereby regulating star formation efficiency \aizw{\citep[e.g.,][]{Hansen2012, Federrath_2013, offner2017,mathew2021} (see also \cite{ frank2014} and \cite{bally2006} for reviews). }

Additionally, such mass ejection could promote material circulation within protostellar disks by lifting dusts from inner disks to outer areas, \aizw{as discussed} in the context of chondrule formation in the early Solar System \aizw{\citep[e.g.,][]{shu1996,shu2001,haugbolle2019}} \aizw{and dust growth 
\citep{tsukamoto2021,Cacciapuoti2024}.} This can be seen as indirect feedback on the disks. However, the more direct and kinematic influence of jets and outflows on protoplanetary disks remains unexplored. 

In this study, we present, for the first time, evidence of a potential direct feedback mechanism of jets on a protoplanetary disk. \aizw{Specifically, we \aizw{identify} an expanding bubble near the star—likely driven by past jet activity—that appears to be interacting with the disk. Although similar bubble morphologies have been observed in systems such as XZ Tau \citep{Krist1997,Krist1999,Kirst2008} and SVS 13 \citep{Hodapp2014} via optical and near-infrared observations, this direct interaction has not been reported previously, perhaps because the bubble's rapid expansion relative to its systemic motion is uniquely evident in the current system.}

\aizw{The present paper is organized as follows.} Section \ref{sec:obs} provides a detailed description for the ALMA observations. Section \ref{sec:results} presents the comprehensive analyses on main three features identified in the system. Section \ref{sec:discuss} presents the discussion, and Section \ref{sec:conclusion} concludes the study. 

\section{Observation and Data Reduction} \label{sec:obs}
\begin{figure*}[t]
\centering
\includegraphics[width=0.99\textwidth]{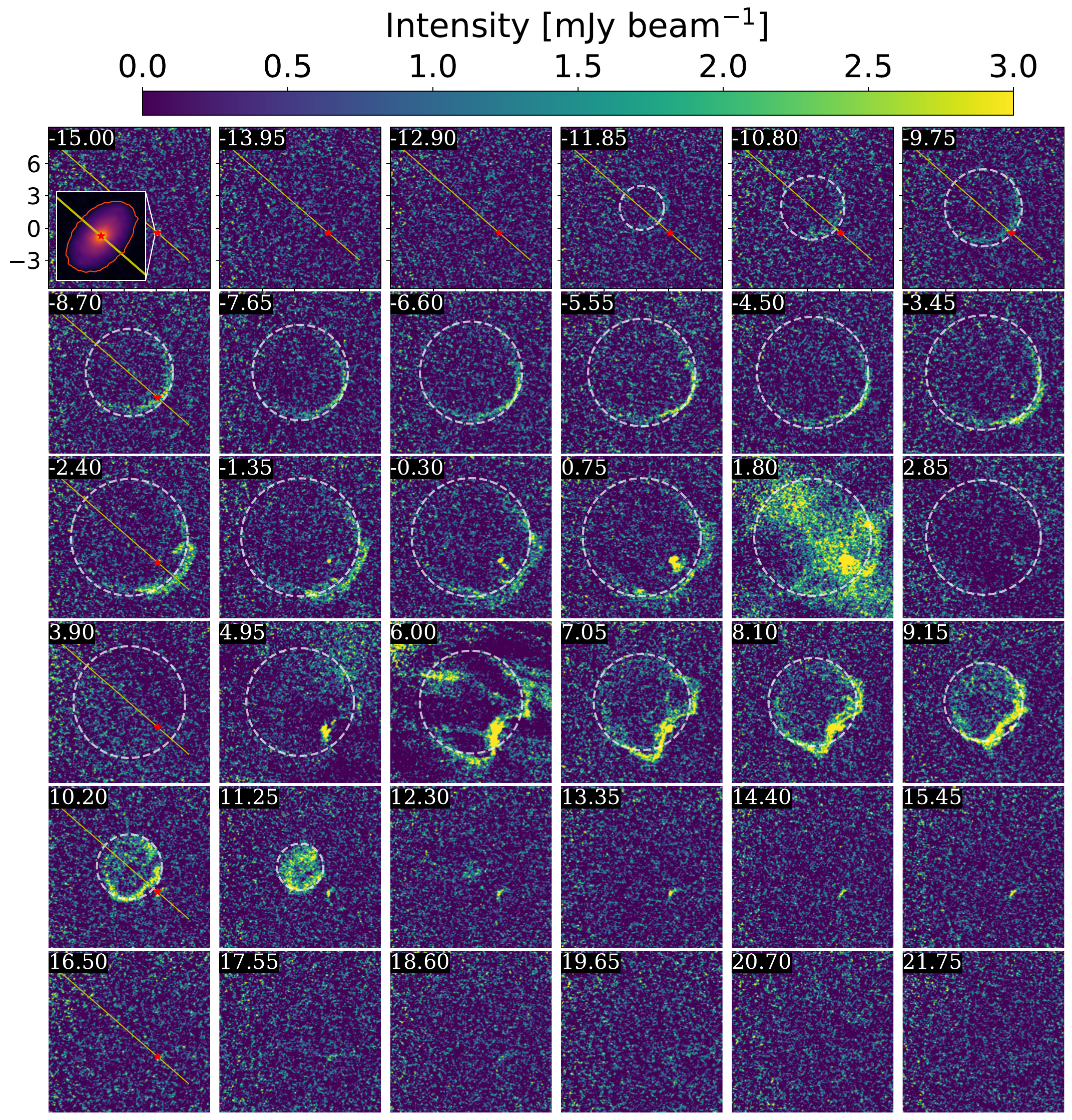}
\caption{Channel maps for $^{12}$CO(2-1) emission of WSB 52. Selected 36 channel maps with velocity spacing of $1.05$~km/s are shown. The line-of-sight velocities are shown in the upper left. The white line in each channel delineates the iso-velocity contour of the expanding bubble model with $(x_{\rm bubble},y_{\rm bubble}, r_{\rm  bubble} ,u_{\rm bubble}, v_{\rm sys, bubble}) = (2.50 \text{ arcsec},1.90 \text{ arcsec},5.5 \text{ arcsec}, 12.5 \text{ km/s}, -0.25 \text{ km/s})$.  \aizw{In the first panel, a zoomed-in view ($0.5\times0.5$ arcsec) in the lower left shows the continuum emission and its contour at a level of 0.1 mJy/beam. The leftmost and uppermost panels highlight the stellar position and the disk axis. The images in $1.05$~km/s $ \leq v_{\rm LOS} \leq$ $4.95$~km/s are affected by cloud contamination.} \aizw{The image is centered on the bubble, with its center being offset from the phase center.} The range of the color bar is limited to 0-3 mJy/beam to enhance the visibility of the expanding bubble. }
\label{fig:channels_maps_all}
\end{figure*}

The target in this paper, WSB 52, is a T Tauri star in Class II phase, located $135.27\pm0.92$~pc away from the Earth \citep{gaia2016,gaia2023}, in the Ophiuchus region. The star harbors a dusty disk with a gap structure, as observed by ALMA in DSHARP (The Disk Substructures at High Angular Resolution Project) \citep{andrews2018}. Table \ref{fig:channels_maps_all} lists the stellar and disk parameters, as used or reported in \cite{andrews2018} and \cite{huang2018}. \aizw{In \cite{andrews2018}, the \aizw{$^{12}\mathrm{CO}\,(J=2\mbox{--}1)$} data for WSB 52 were described as ``cloud (severe), complex outflow" in their Table 5. While we identify the noted cloud contamination, our analysis provides further insight into the system's nature. Note that, aside from WSB 52, } we also examined the other \aizw{DSHARP} targets but did not identify similar events in any of the other systems.

We began our analysis by examining  the \aizw{$^{12}\mathrm{CO}\,(J=2\mbox{--}1)$} cube data, available on the DSHARP data release webpage.  However, we found that the velocity coverage of the released image cube was insufficient to fully capture the features of our interest; therefore, we reconstruct the image cube from the raw data, extending the line-of-sight velocity range to $v_{\rm LSR} = -15.70$~km/s to $29.45$~km/s. The measurement set, downloaded from the ALMA archive, is processed using the Common Astronomy Software Applications (CASA) pipeline (version 4.7.2) by the East Asian ALMA Regional Center (EA-ARC), followed by self-calibration with the standard DSHARP analysis script using CASA 6.6.5. For CLEAN imaging, we basically adopt the same parameters as the standard script but slightly adjust the following parameters; ${\tt gain}=0.2$, ${\tt niter}=200000$, ${\tt threshold}=2.5$~mJy, and ${\tt imsize}= 2000$. We apply the primary beam correction to the image.

The reconstructed cube data have a \aizw{channel spacing} of $0.35$ km/s and an image pixel size of $0.01$~arcsec. \aizw{As noted in \cite{andrews2018}, the actual velocity resolution is about two channels, because of Hanning smoothing in the ALMA correlators.} The synthesized beam size is $0.141 \times 0.092$~arcsec with a position angle of $-85.5^{\circ}$. \aizw{As \cite{andrews2018} described this target as ``cloud (severe), complex outflow"}, contamination from extended molecular cloud emission around WSB 52 obscured the circumstellar gas, including disk emission near the systemic velocity ($v_{\text{sys},\star} = 3.9$~km/s, as determined in Sec \ref{sec:deformed}) within the range $v_{\rm LSR} = 2.85$ to $4.6$~km/s. Nonetheless, our focus on velocity components with minimal cloud contamination ensures that this issue does not compromise our primary findings.

\begin{table*}
\begin{center}
\caption{Stellar and Disk Parameters Used in This Study \label{table:parameters}}
\begin{tabular}{lll}
\hline \hline
Parameter  &  Value  & Reference  \\
\hline 
$M_{\star}$ [$M_{\rm sun}$] & 0.4786 & \cite{andrews2018}\\
Distance [pc] & 135.27  &  \cite{gaia2023}\\
$x_{\star}^{\dagger}$ [arcsec] & $-$0.12 &  \cite{huang2018}\\
$y_{\star}^{\dagger}$ [arcsec] & $-$0.43 &  \cite{huang2018}\\
$i$ [deg] & 54.4 &  \cite{huang2018}\\
PA$^{\dagger\dagger}$ [deg] & 138.4&  \cite{huang2018}\\
\hline
\label{S3_parametric_study}
\end{tabular}\\
$^{\dagger}$ $(x, y) = (\Delta \text{RA}, \Delta \text{Dec})$ measured from the observational center. \\ 
$^{\dagger\dagger}$ Position angle for the disk major axis measured from the north direction.
\end{center}
\end{table*}

\begin{deluxetable*}{ll}
\tablecaption{Model Parameters\label{table:model_parameters}}
\tablewidth{0pt}
\tablehead{
\colhead{} & \colhead{\textbf{Value}}
}
\startdata
\textbf{Expanding Bubble Model in Sec \ref{sec:exp_bubble}}\\
\hline 
$x_{\rm bubble}$ [arcsec] & 2.50  \\
$y_{\rm bubble}$ [arcsec] & 1.90 \\
$r_{\rm bubble}$ [arcsec] & 5.5  \\
$v_{\rm sys,bubble}$ [km/s] & $-$0.25 \\
$u_{\rm bubble}$ [km/s] & 12.5 \\
\hline 
\textbf{Shock Boundary Model in Sec \ref{sec:shock_boundary}}\\
\hline 
$r_{\rm bubble}$ [arcsec] & 5.5 / 6.5  \\
$a$ & 2.14 \\
$h_{\rm shock}$ [arcsec] & 0.18 \\
\aizw{$\zeta_0$} [arcsec] & $-$0.28 \\
\hline 
\textbf{Keplerian Disk Model in Sec \ref{sec:deformed}} \\
\hline 
$v_{\text{sys},\star}$ [km/s] & 3.9 \\
$h_{\rm disk}$ [au] & 5 \\
$p$ & 1 \\
\enddata
\end{deluxetable*}

\section{Result} \label{sec:results}
Figure \ref{fig:channels_maps_all} presents the selected channel maps with velocity spacing of $1.05$~km/s for the $^{12}$CO emission. After a thorough inspection of the channel maps, three distinct features have been identified as follows: (1) shell-like patterns in $v_{\mathrm{LSR}} = -11.85$ to $0.75$ km/s and $v_{\mathrm{LSR}} =4.95$ to $11.25$ km/s in the figure, sharing the same central position offset from the star, (2) a concave morphology at $v_{\mathrm{LSR}} = 4.95$-$10.20$ km/s in the southwestern part of the shells, and (3) spatially compact emissions in the vicinity of the star at $v_{\mathrm{LSR}} = -6.60$ to $18.60$ km/s. These features can be interpreted as follows: (1) an expanding bubble offset from the stellar position, (2) a shock boundary between the bubble and the star's vicinity, and (3) a deformed protoplanetary disk. 

\begin{figure*}[t]
\centering
\includegraphics[width=.8\textwidth]{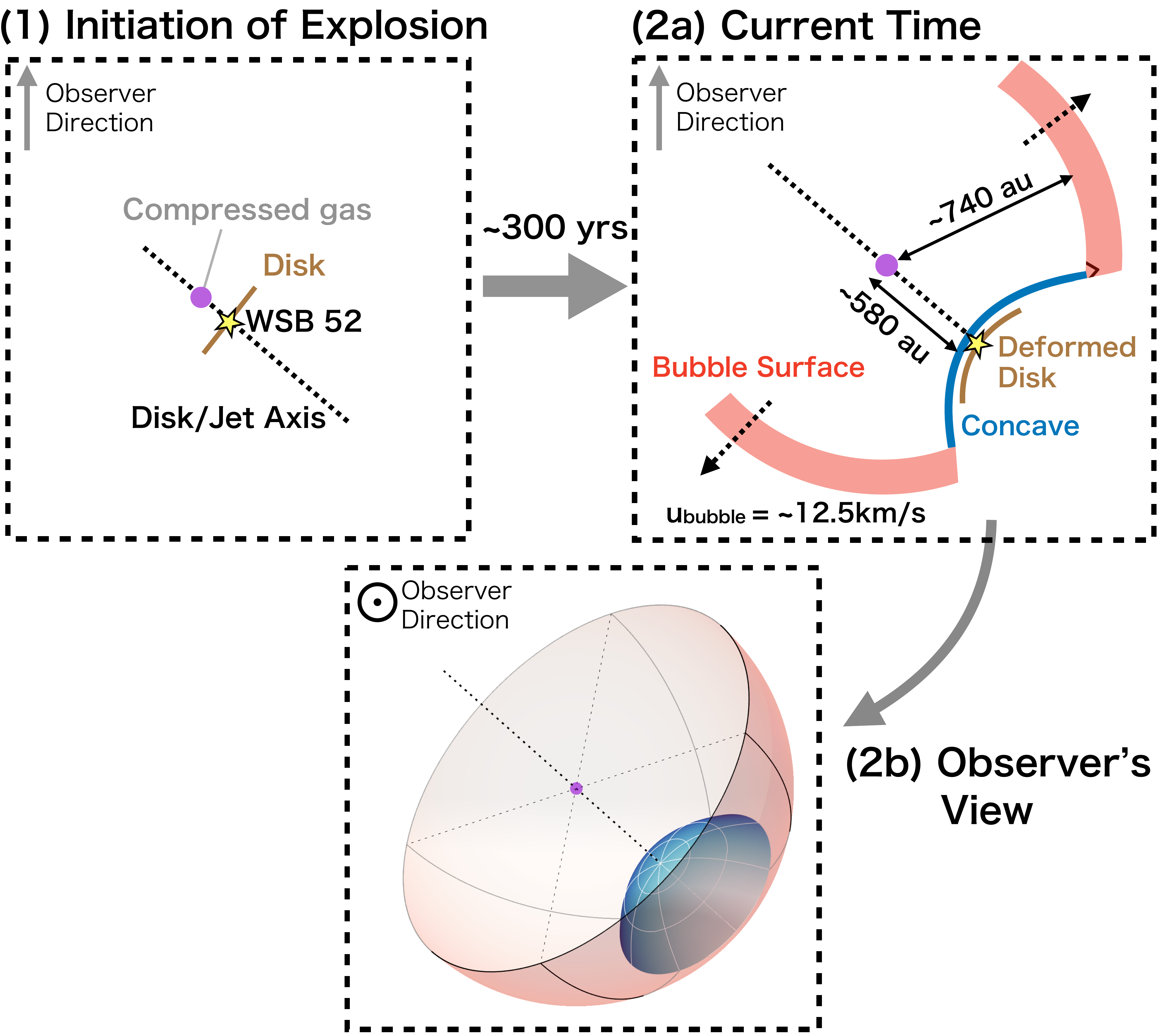} 
\caption{ Schematic illustration of proposed models explaining data. The upper left panel shows the illustration before the initiation of the explosion. The upper right panel illustrates the current state of the system, and the lower panel presents the observed view. The configuration of the bubble and the shock boundary in panel (2b)  corresponds to that in Figure \ref{fig:xi_coord}.  } 
\label{fig:summary}
\end{figure*}

\aizw{Figure \ref{fig:summary} shows the proposed models that will be discussed in detail throughout the paper; the bubble is likely driven by jet compression, which sweeps up the circumstellar material and interacts with the disks. Panel (2a) illustrates the current state of the system as inferred from the observations.} In the remaining of this section, we analyze the structures identified \aizw{in the observations.}

\subsection{The expanding bubble} \label{sec:exp_bubble}
Figure \ref{fig:channels_maps_all} shows shell-like patterns that vary in size with velocity, ultimately converging to a single point. This morphology appears to be explained by a uniformly expanding bubble. \aizw{The bubble is not perfectly spherical but shows notable asymmetry, appearing particularly bright near the star. This point is discussed in detailed in Section \ref{sec:scenario}.}

To constrain the physical property of the bubble, we develop a simple model, which is characterized by the central position $(x_{\rm bubble}, y_{\rm bubble})$, bubble radius $r_{\rm bubble}$, radial expansion velocity $u_{\rm bubble}$, and systematic velocity $v_{\rm sys, bubble}$. In the modeling, we adopt a Cartesian coordinate system $(x, y, z)$. We assume $(x, y) = (\Delta \text{RA}, \Delta \text{Dec})$, where $\Delta \text{RA}$ and $\Delta \text{Dec}$ are the offsets in right ascension and declination, respectively, from the phase center of the image. The $z$-axis represents the relative depth to the bubble center; we assume that the bubble center is located at $(x_{\rm bubble}, y_{\rm bubble}, 0)$, and a positive $z$-value represents the side that is farther away from the observer. We here use arcseconds as the unit for the scale of the coordinate system, but we can convert it into physical scales by using the distance $d=135.27$ pc. 

In the rest frame of the bubble, the velocity field is defined as: 
\begin{eqnarray}
    \mathbf{u}(\mathbf{r}) &=&  u_{\rm bubble} \left(\frac{\mathbf{r}}{r_{\rm bubble}}\right),  \label{eq:bubble_vel}
\end{eqnarray}
where ${\bf r} = (x - x_{\rm bubble}, y- y_{\rm bubble}, z)$ denotes a relative vector from the bubble center to the point on the sphere. 

We assume the line-of-sight systematic velocity of the bubble to be $v_{\rm sys, bubble}$. The line-of-sight radial velocity for the expanding bubble obtained by an  observer is then given as follows:
\begin{eqnarray}
    v_{\rm LOS}(z) &=& v_{\rm sys, bubble} + u_{\rm bubble}\left(\frac{z}{r_{\rm bubble}}\right). 
\end{eqnarray}
Conversely, for a given constant observed velocity $v_{\rm LOS}$, the cross-section of the sphere is characterized by the depth $z(v_{\rm LOS})$ and the radius $r(v_{\rm LOS})$ as follows: 
\begin{eqnarray}
    z(v_{\rm LOS}) &=& r_{\rm bubble} \left(\frac{v_{\rm LOS} - v_{\rm sys,  bubble}}{u_{\rm bubble}}  \right),  \\
    r(v_{\rm LOS}) &=& r_{\rm bubble} \sqrt{1 - \frac{z^{2}(v_{\rm LOS})}{r_{\rm bubble}^{2}}}.
\end{eqnarray}

The model described above provides the location of the bubble rim for each observed line-of-sight velocity. \aizw{We have opted to employ a visual optimization of the parameters by comparing the observed bubble rims to the model prediction. This approach is motivated by the fact that  a numerical optimization is susceptible to the arbitrary choices of the optimization function and criteria when analyzing such a complex phenomenon.} Specifically, given the parameters and the array of the observed velocities, we plot the circles with the center $(x_{\rm bubble}, y_{\rm bubble})$ and the radii of $r(v_{\rm LOS})$, and then compare them to the cube data. \aizw{The obtained solution is in strong agreement with the observed bubble, thereby satisfying the objective of this analysis to characterize the emission features in a coherent manner.}

\aizw{We overlay the continuum emission contour and the disk axis in Figure \ref{fig:channels_maps_all}. The image is taken from the DSHARP observation at \url{https://almascience.eso.org/almadata/lp/DSHARP/}, and a contour level of 1~mJy/beam is used to approximate the outer disk radius. Both the bubble and the concave structure appear symmetric with respect to the disk axis, also as discussed in the next subsection. Consequently, the disk axis seems to point toward the bubble center, not only in the sky plane but also in three-dimensional space. We thus assume that} the bubble center to be on the line of the disk axis \aizw{in determining $(x_{\rm bubble}, y_{\rm bubble})$.} 

Table \ref{table:model_parameters} presents our optimized parameters. White lines in Figure \ref{fig:channels_maps_all} delineate the bubble rims in our model. Despite its simplicity, the model demonstrates a high degree of agreement with the observational data. The inferred radius of the bubble is approximately 750~au, and the expansion velocity is 12.5 km/s. \aizw{The center of the bubble is derived as ($x_{\rm bubble}, y_{\rm bubble}$) = ($2.50", 1.90"$), indicating a distance between the bubble center and the star of 580 au if $i = 58.4^{\circ}$. The stellar location is within the bubble, thereby suggesting an interaction between circumstellar materials and the bubble. In addition, $v_\mathrm{{sys, bubble}}$, $-0.25$ km/s, is blue-shifted from the systemic velocity of the star.} 

We further estimate the bubble mass and kinetic energy using $^{12}$CO(2-1) intensities under the assumption that the emission is not completely optically thick. In the  analysis, we mask out the disk gas emission within a 1.5~arcsec radius around the star to isolate the shell. The detail of the  calculation is given in Appendix \ref{sec:mass_energy_bubble}.  Assuming local thermodynamic equilibrium (LTE) with gas temperatures ranging from 25~K to 100~K, we derive a total mass of $M_{\rm shell} = (0.2-1.1) \times10^{-4}$~$M_{\odot}$.  Combined with the mass and the expansion velocity with 12.5 km/s, we estimate a kinetic energy of $E_{\rm shell} = (0.3-1.6) \times10^{41}$~erg. As discussed in Section \ref{sec:discuss}, this energy scale is consistent with that of the jets. 

\subsection{The shock boundary between bubble and the stellar vicinity}  \label{sec:shock_boundary}

\begin{figure*}[t]
\centering
\includegraphics[width=0.43\textwidth]{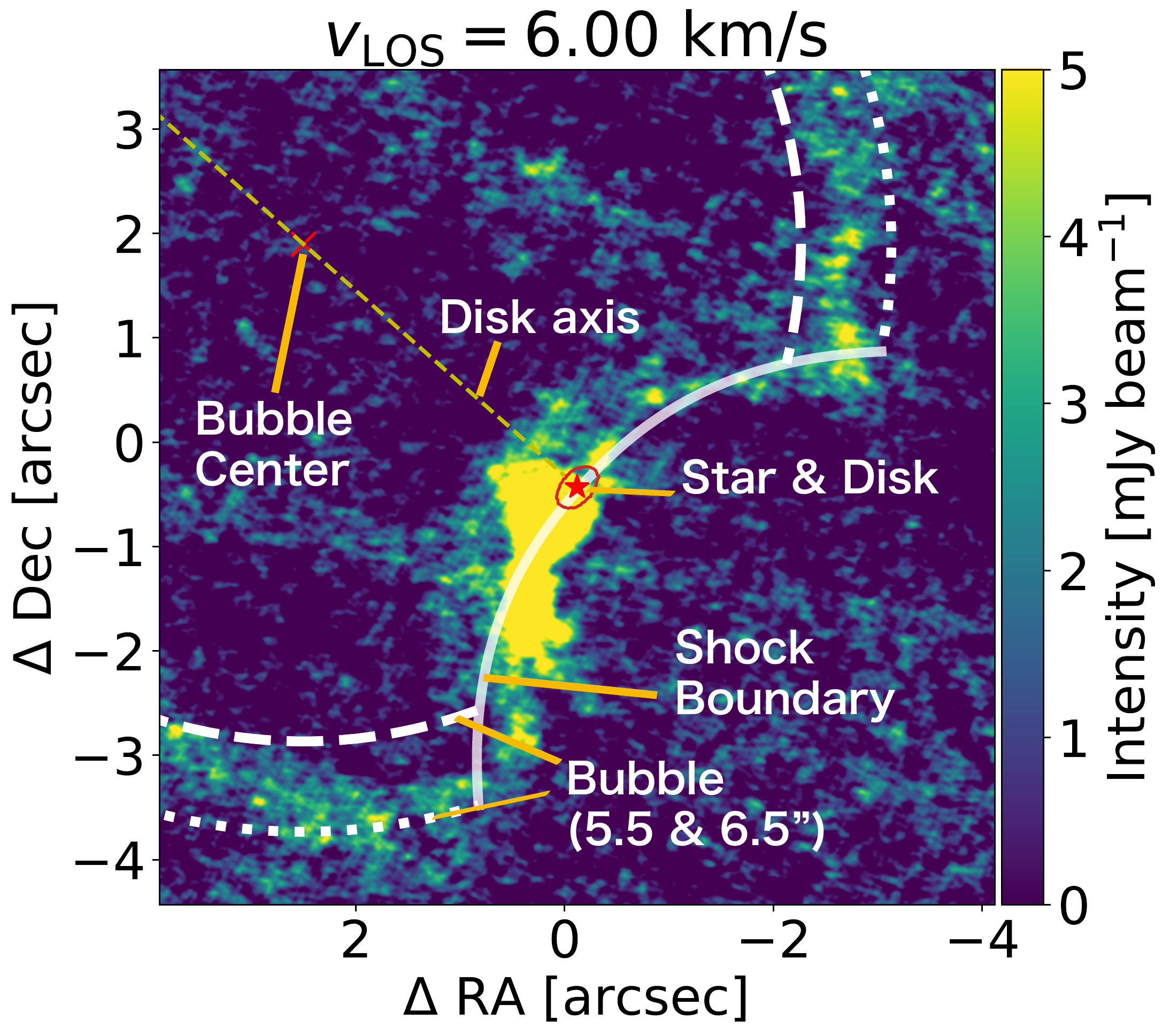}
\includegraphics[width=0.37\textwidth]{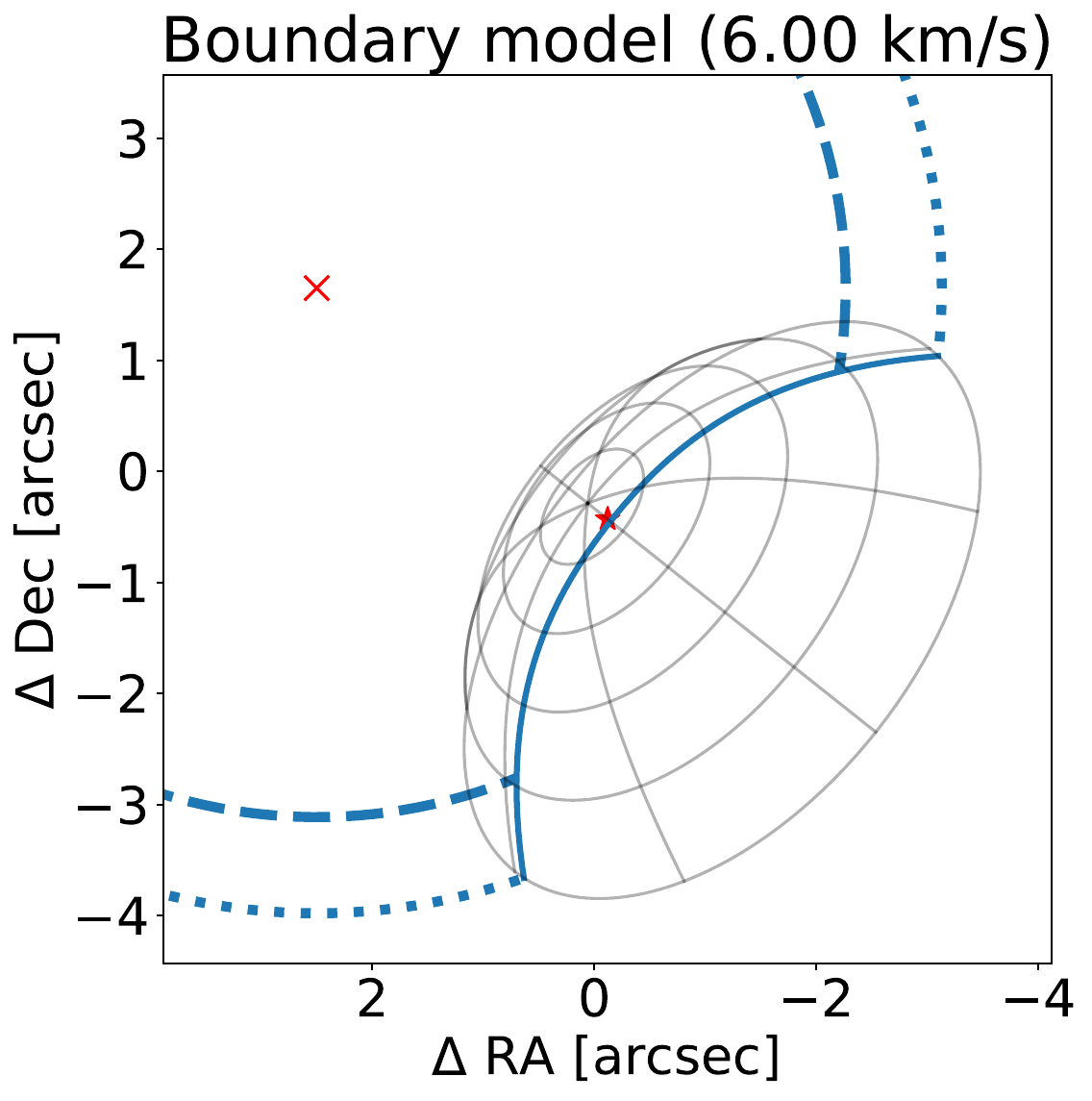}
\includegraphics[width=0.43\textwidth]{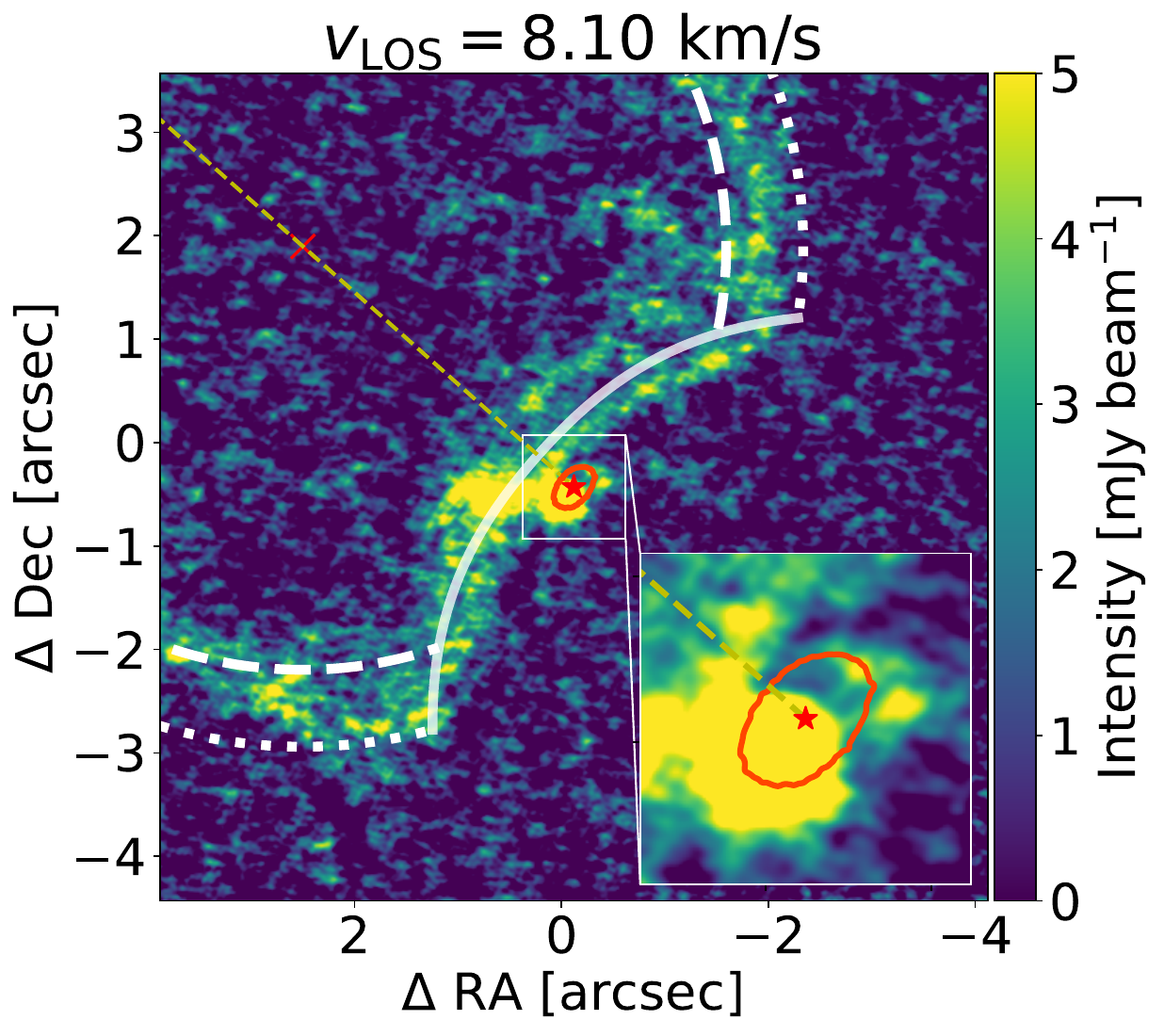}
\includegraphics[width=0.37\textwidth]{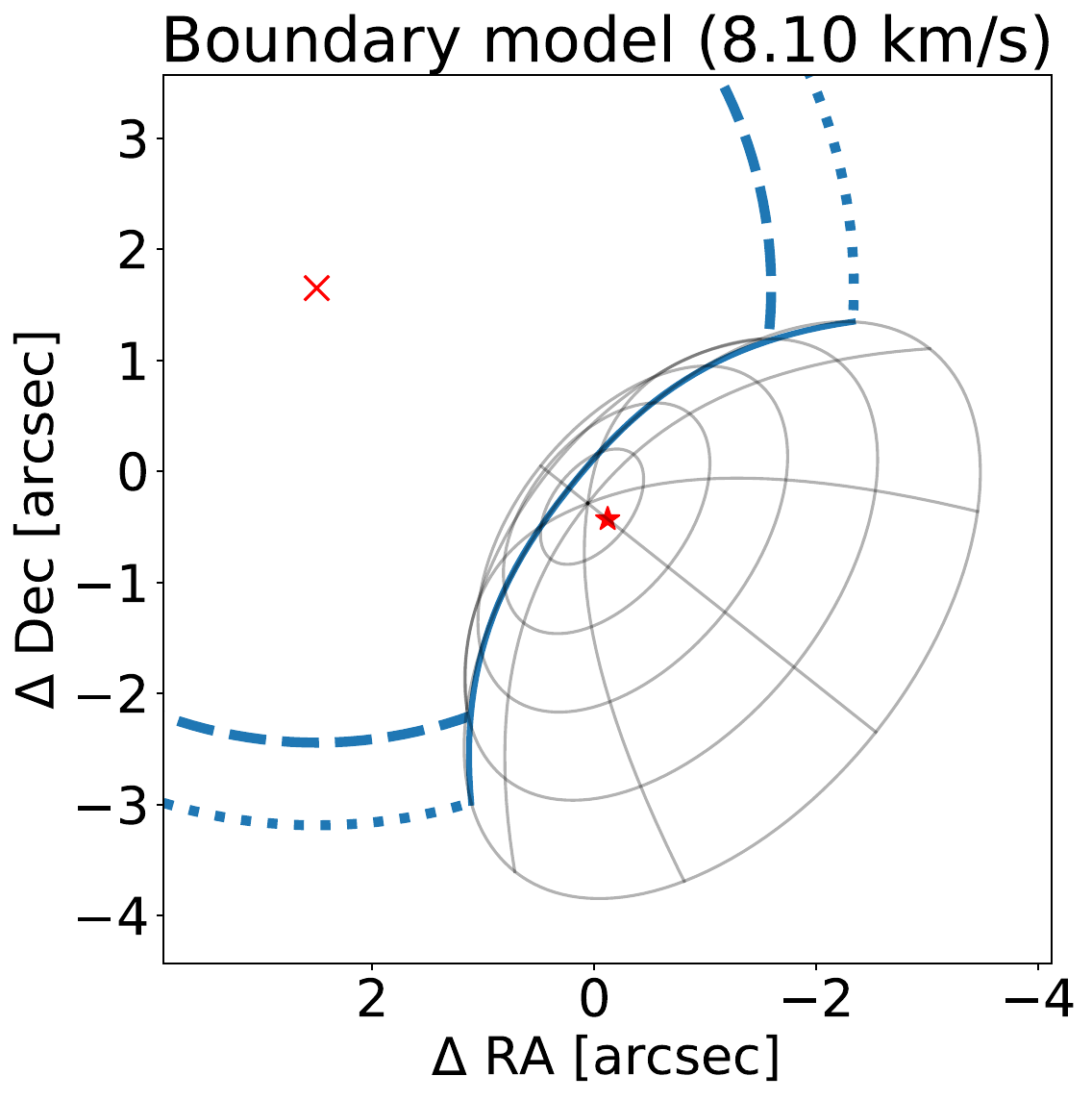}
\includegraphics[width=0.43\textwidth]{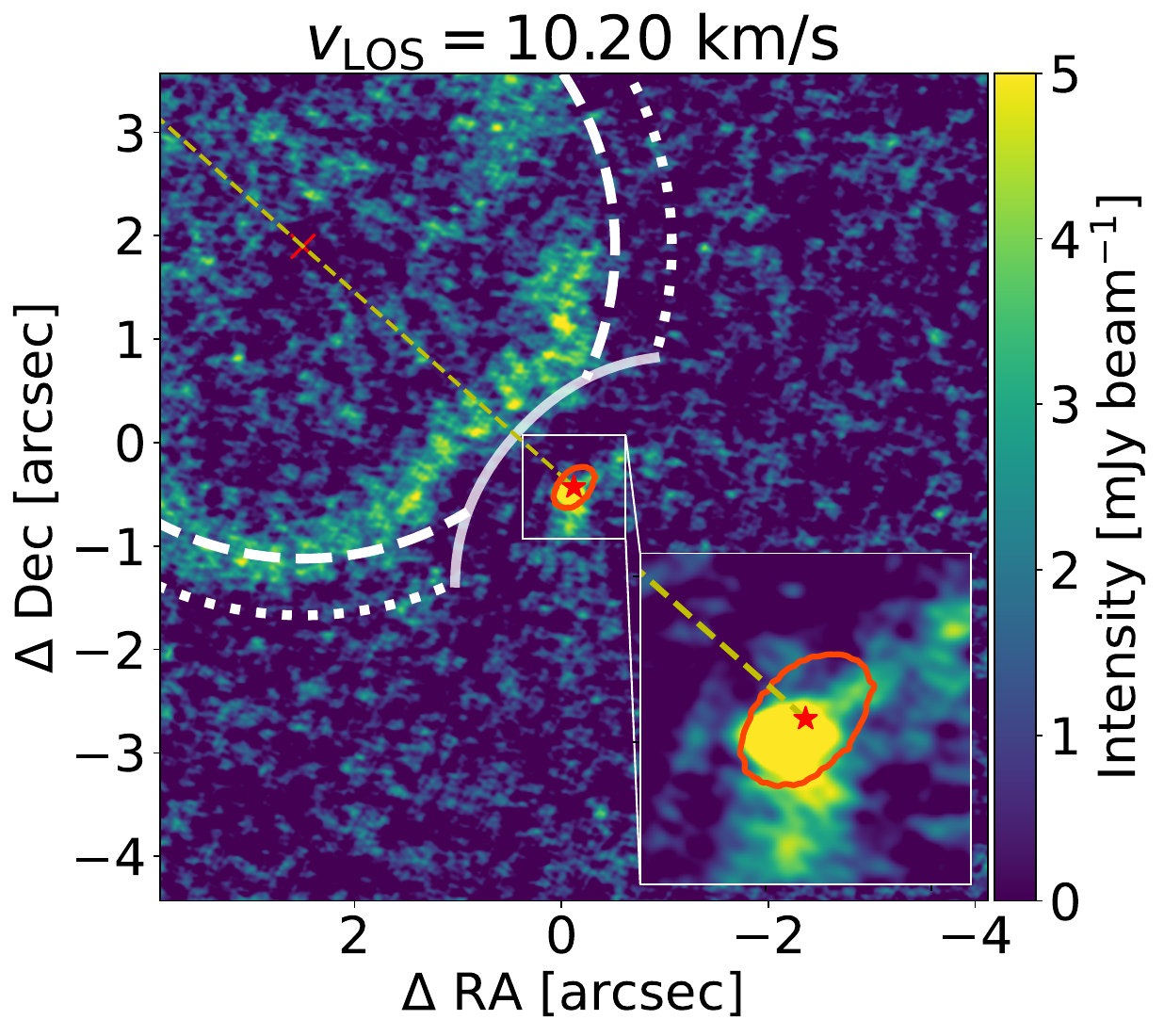}
\includegraphics[width=0.37\textwidth]{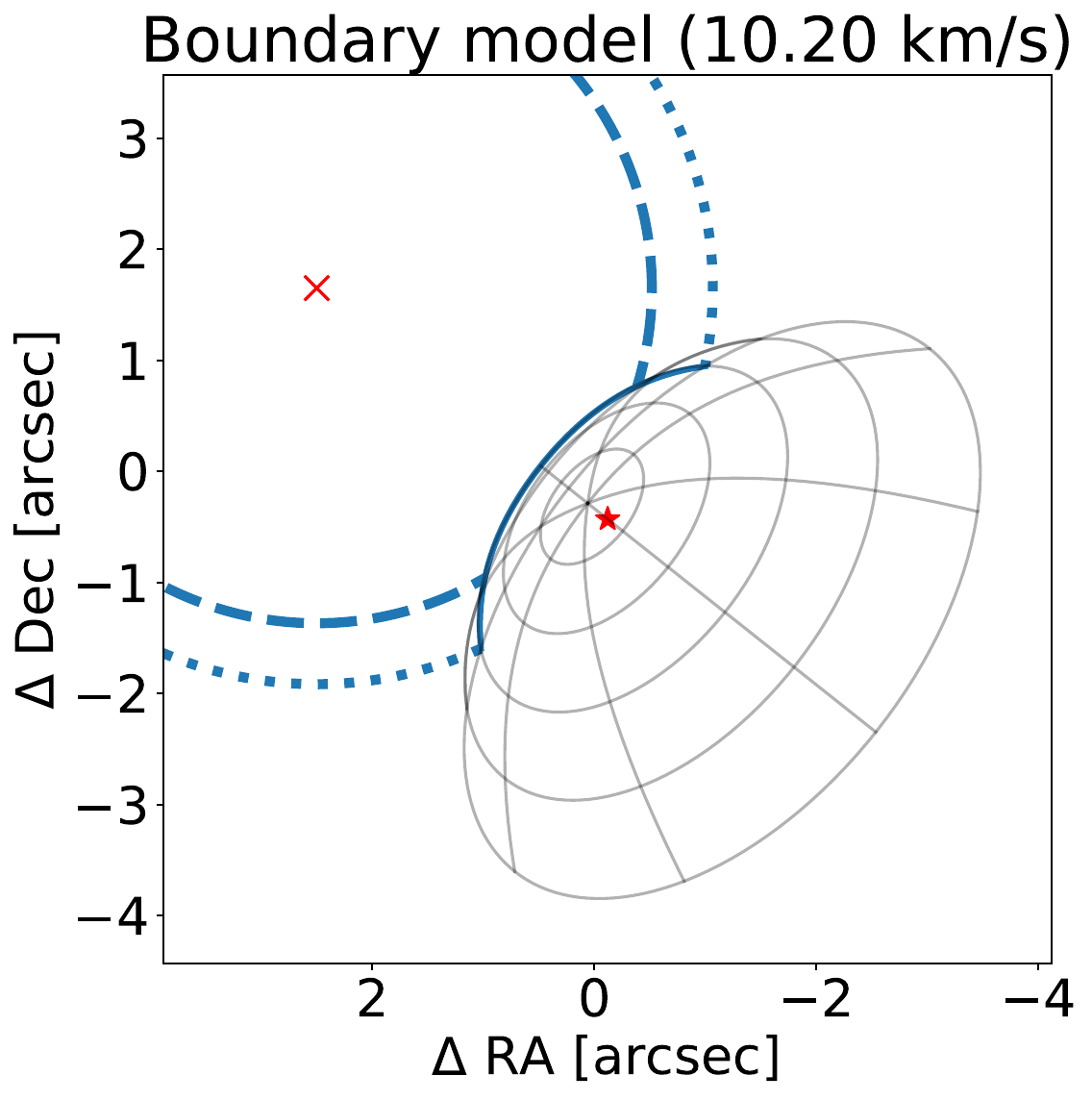}
\caption{ Close-up view of the shock boundary between the bubble and the star. Zoomed-in views of three selected channel maps from Figure \ref{fig:channels_maps_all} are shown in the left columns. The iso-velocity contours for the bubble and the boundary models are also illustrated. The spatial morphology of the shock boundary is presented in the right panels. \aizw{Zoomed-in views of the continuum emission contour at a level of 0.1~mJy/beam are presented in the two panels.} The range of the color bar is limited to 0-5 mJy/beam.
}  
\label{fig:channels}
\end{figure*}

As in Figure \ref{fig:channels_maps_all}, the bubble is not perfectly spherical; it exhibits a concave morphology near the star. Figure \ref{fig:channels} presents a zoomed-in view of this feature across selected velocity channels. \aizw{We overplot the continuum emission contour at a level of 0.1~mJy/beam in Figure \ref{fig:channels}. }

Notably, the disk axis points toward the bubble center, and the shock boundary appears to be symmetric along this disk axis as well. These observations prompt two assumptions: (a) the shock boundary is nearly axisymmetric with respect to the disk axis, and (b) the bubble center lies on the disk axis in three-dimensional space, \aizw{as described in Section \ref{sec:exp_bubble}.} Under the assumption (b), we estimate the star's line-of-sight depth relative to the bubble center by using the disk inclination and the projected distance on the sky. Specifically, for the stellar position $(x_{\star}, y_{\star}, z_{\star})$, where $x_{\star}, y_{\star}$ are given in Table \ref{table:parameters}, we determine $z_{\star}$ as follows:
\begin{equation}
    z_{\star} = \frac{\sqrt{(x_{\star} -x_{\rm bubble})^{2} + (y_{\star} -y_{\rm bubble})^{2}} }{\tan i}, 
\end{equation}
where $i$ is the disk inclination. 

\begin{figure*}[t]
\centering
\includegraphics[width=0.50\textwidth]{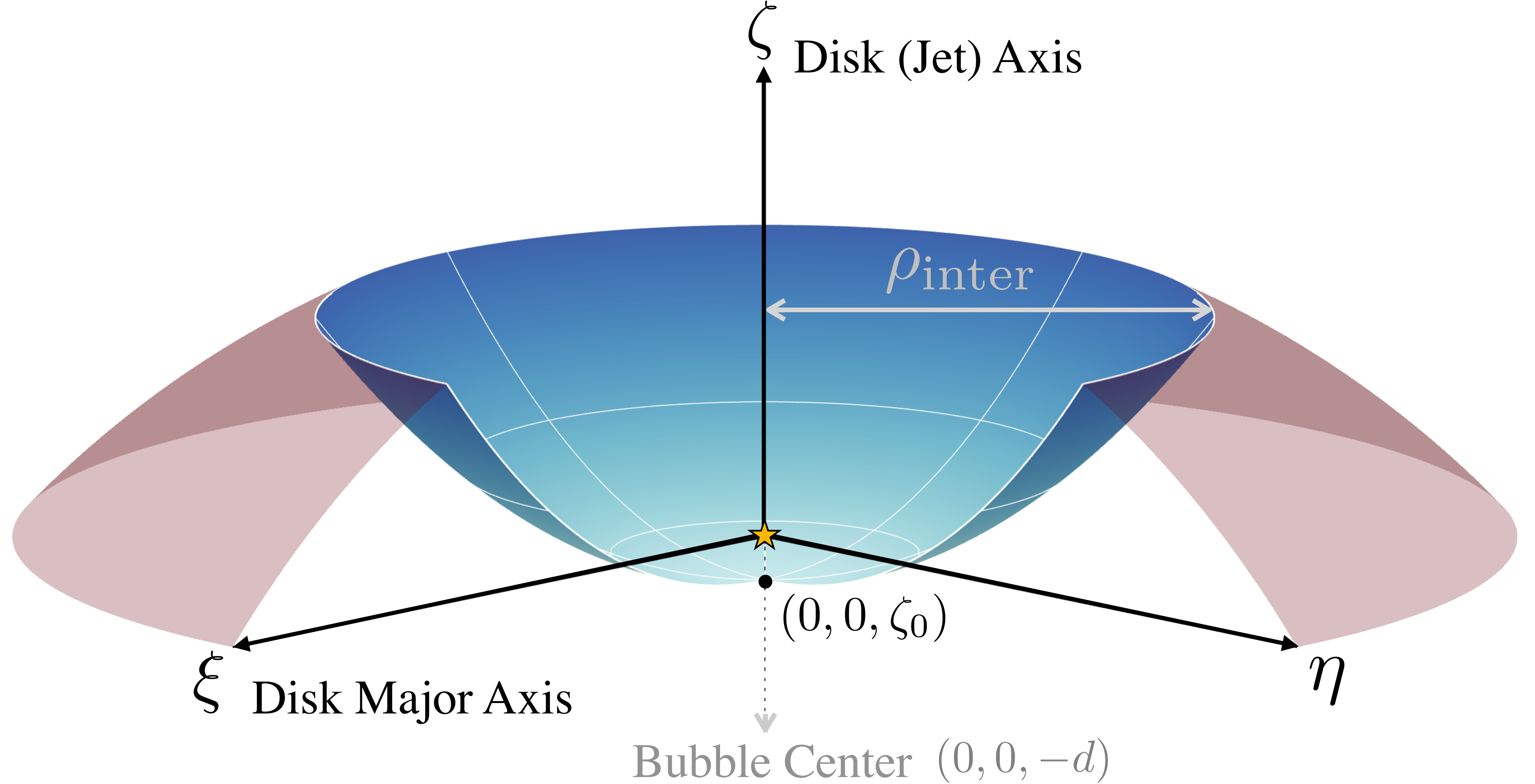}
\includegraphics[width=0.40\textwidth]{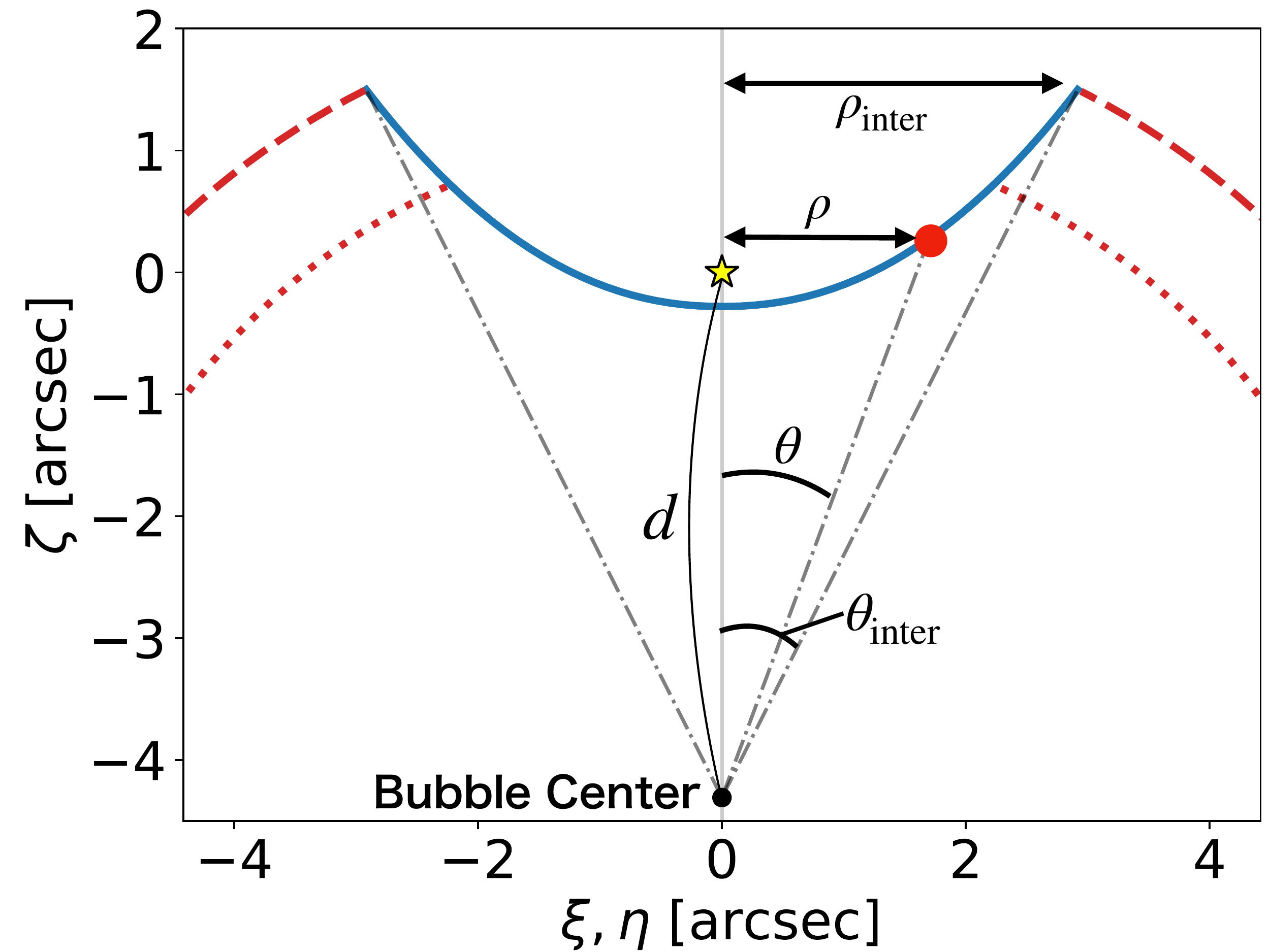}
\caption{(left) Illustration of the $(\xi, \eta, \zeta)$ coordinate system used to define the shock boundary model. (right) Side view of the shock boundary model. The parameters in Table \ref{table:model_parameters} are assumed. The blue line represents the shock boundary, while the red dotted and dashed lines correspond to bubbles with radii of 5.5 and 6.5$\arcsec$, respectively. The star symbol at the origin indicates the stellar location, and the black point corresponds to the bubble center. The adopted values for $\rho_{\rm inter}$ and $\theta_{\rm inter}$ are shown assuming $r_{\rm bubble}=6.5 \arcsec$. }
\label{fig:xi_coord}
\end{figure*}

To interpret the concave morphology, we develop a simple analytical model. We introduce a coordinate system $(\xi, \eta, \zeta)$ as in the left panel of Figure \ref{fig:xi_coord}. The $\zeta$-axis is assumed to pass through both the star and the center of the bubble, with the positive direction of the axis pointing from the bubble center toward the star, and the origin is taken to be the stellar position. The $(\xi, \eta)$-plane corresponds to the disk plane, and the $\xi$-axis aligns with the disk major axis in the sky plane. We denote the radial distance from the $\zeta$ axis by $\rho$, as shown in the right panel of Figure \ref{fig:xi_coord}, as follows: 
\begin{equation}
\rho = \sqrt{\xi^2 + \eta^2}. 
\end{equation}

We model the vertical position of the shock boundary using a simple power-law form with a vertical offset:
\begin{eqnarray}
\zeta_{\rm surf} (\rho) = h_{\rm shock} \left(\frac{\rho}{1 \; \text{arcsec}} \right)^{a}  + \zeta_{0}, 
\end{eqnarray}
where $h_{\rm shock}$ determines the height of the surface, $a$ determines the steepness of the surface, and $\zeta_0$ determines the vertical offset from the stellar position. The model is axisymmetric with respect to the $\zeta$ axis. 

In the $(\xi, \eta, \zeta)$ coordinate system, the bubble center is located at $(0, 0, -d)$, where $d$ is the distance between 
the star and bubble center, 
\begin{equation}
 d = \sqrt{(x_{\star} -x_{\rm bubble})^{2} + (y_{\star} -y_{\rm bubble})^{2}  + z_{\star} ^{2}}. 
\end{equation}

We consider two cases for our models; the bubble model and the shock boundary model. To distinguish between the two cases, we  introduce the angle $\theta$, which is measured from the $\zeta$-axis within a spherical coordinate system that has its origin at the bubble center $(0,0,-d)$. The angle $\theta$ is depicted in the right panel of Figure \ref{fig:xi_coord}: 

\begin{equation}
\tan \theta = \frac{\rho}{d+\zeta},
\end{equation}
or 
\begin{equation}
    \theta = \arctan \left(\frac{\rho}{d+\zeta}\right).
\end{equation}

We categorize the points on the surface  (indicated by the red point in the panel of Figure \ref{fig:xi_coord}) into two cases based on the angle $\theta$ as follows: 
\[
\begin{cases} 
\text{Shock boundary model} & \text{if } \theta < \theta_{\rm inter} \\
\text{Bubble model} & \text{if } \theta \geq \theta_{\rm inter},  
\end{cases}
\]
where $\theta_{\rm inter}$ is the boundary value, as depicted in the right panel of Figure \ref{fig:xi_coord}.

The intersection between the bubble and the shock boundary is characterized by $\theta_{\rm inter}$ or $\rho=\rho_{\rm inter}$, which can be determined by solving following equation: 
\begin{equation}
r_{\rm bubble}^{2} = \rho_{\rm inter}^{2} + (\zeta_{\rm surf}(\rho_{\rm inter}) + d)^{2},
\end{equation} 
where both the left-hand side and the right-hand side represent the square of the distance from the bubble center. The polar angle $\theta_{\rm inter}$ is derived from $\rho_{\rm inter}$ as follows 
\begin{equation}
 \theta_{\rm inter} =  \arctan\left(\frac{\rho_{\rm inter}}{d + \zeta_{\rm surf}(\rho_{\rm inter})}\right). 
\end{equation}
Note that the values of $ \theta_{\rm inter}$ and $\rho_{\rm inter}$ depend on $r_{\rm bubble}$.

For the bubble model, the spatial distribution and velocity fields are taken to be same as those in the simple expansion model in Sec \ref{sec:exp_bubble}. In the shock boundary model, we assume that the velocity direction aligns with the radial direction from the bubble center, and the velocity amplitude is constant, $u_{\rm bubble}$. Specifically, the velocity field on the boundary $\mathbf{u}$ in the rest frame of the bubble center is given as follows: 
\begin{equation}
\mathbf{u} = u_{\rm bubble} \left(\frac{\mathbf{ S}}{||\mathbf{ S}||}\right) , 
\end{equation}
where $\mathbf{S}$ is the position vector from the bubble center to the point on the boundary. This equation is assumed in a similar manner to Eq (\ref{eq:bubble_vel}).

Assuming the above models, we derive analytical expressions for the predicted shock-boundary contours with fixed $v_{\rm LOS}$, as presented in Appendix \ref{sec:analytical_shock}. We optimize the shell models by comparing the predicted contours $(x_{\rm contour}(v_{\rm LOS}, \rho), y_{\rm contour}(v_{\rm LOS}, \rho))$ with the observed shock boundaries. \aizw{We initially attempted manual optimization as in Section \ref{sec:exp_bubble}; however, due to the model's increased complexity, we relied on numerical optimization.} To facilitate efficient optimization, we manually approximate the observed shock boundaries by discrete points along each channel. For each velocity, we calculate the sum of the distances from the points to the predicted model contours. The total distance metric is defined as the sum of these minimum distances across all selected channels. We minimize this aggregate distance by adjusting the parameters of the bubble model. 

Table \ref{table:model_parameters} presents the derived parameters for the bubble model. To account for the range in bubble width, we adopt two values for $r_{\rm bubble}$, namely $5.5 \arcsec$ and $6.5\arcsec$. The optimized models are shown as white dotted and dashed lines in Figure \ref{fig:channels}. Our model reasonably replicates the observed shock boundary morphology, supporting our estimation of the stellar position. \aizw{The three-dimensional distance between the shock's apex and the stellar position is $\zeta_0 = -0.28\arcsec$, which corresponds to approximately 38 au.} The right panel in Figure \ref{fig:xi_coord} shows a side view of the optimized shock boundary models with $r_{\rm bubble}=5.5 \arcsec$ and $6.5\arcsec$\aizw{, where $(\rho_{\rm inter}, \theta_{\rm inter}) = (2.23\arcsec, 23.9^{\circ})$ and $(2.92\arcsec, 26.7^{\circ})$, respectively. }


\subsection{The deformation of the disk by the expanding bubble}
\label{sec:deformed}

\begin{figure*}[t]
\centering 

\includegraphics[width=0.32\textwidth]{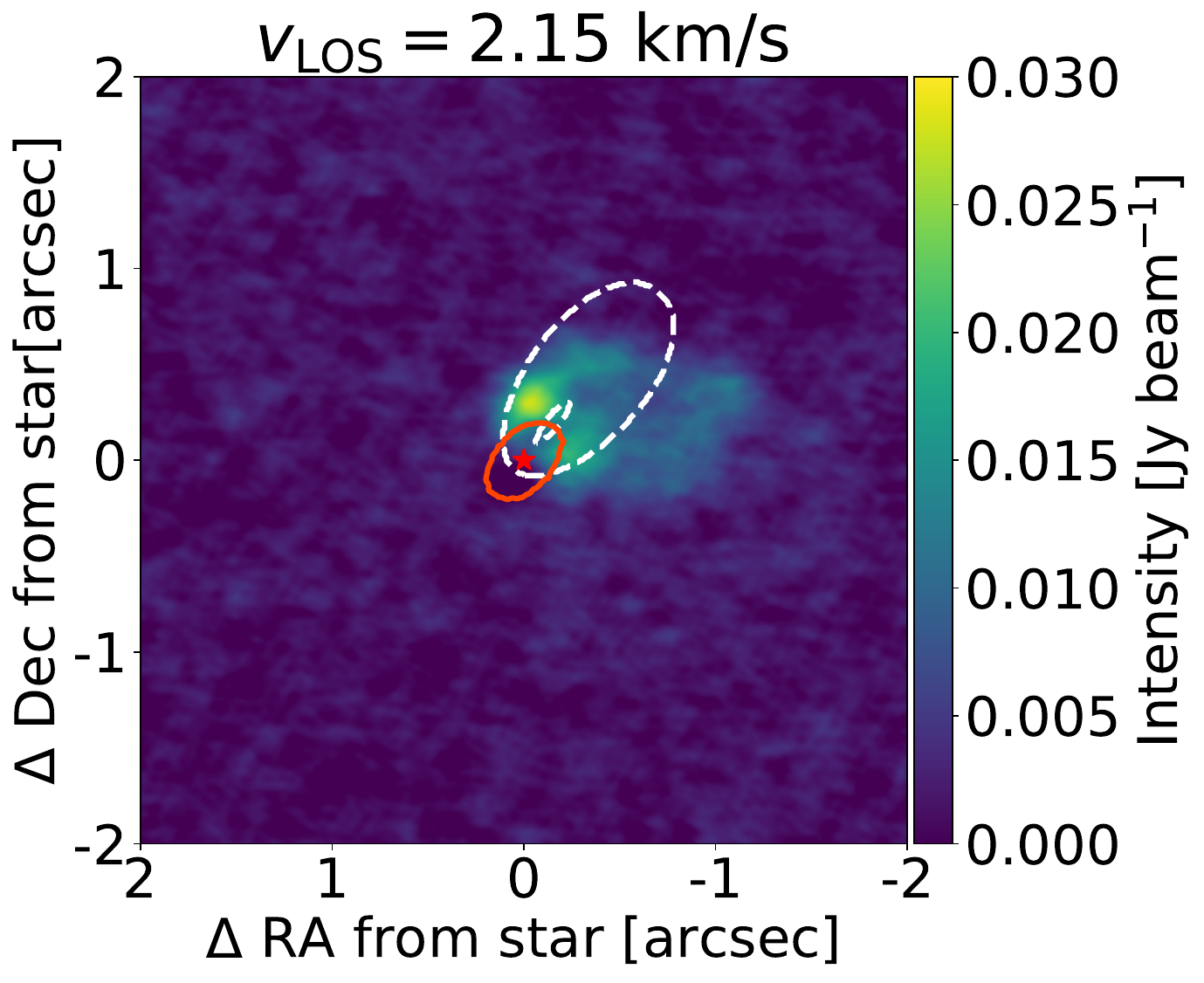}
\includegraphics[width=0.32\textwidth]{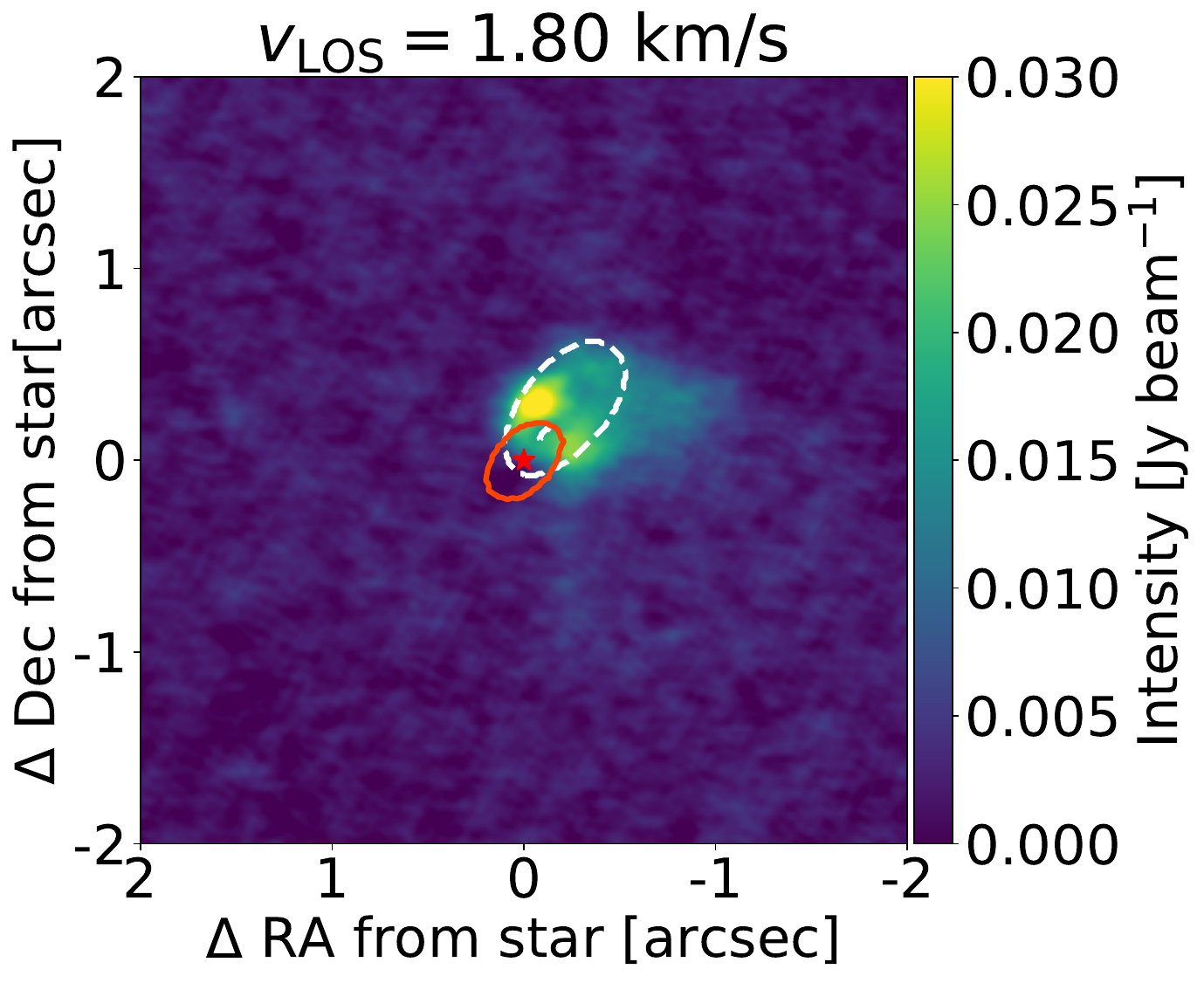}
\includegraphics[width=0.32\textwidth]{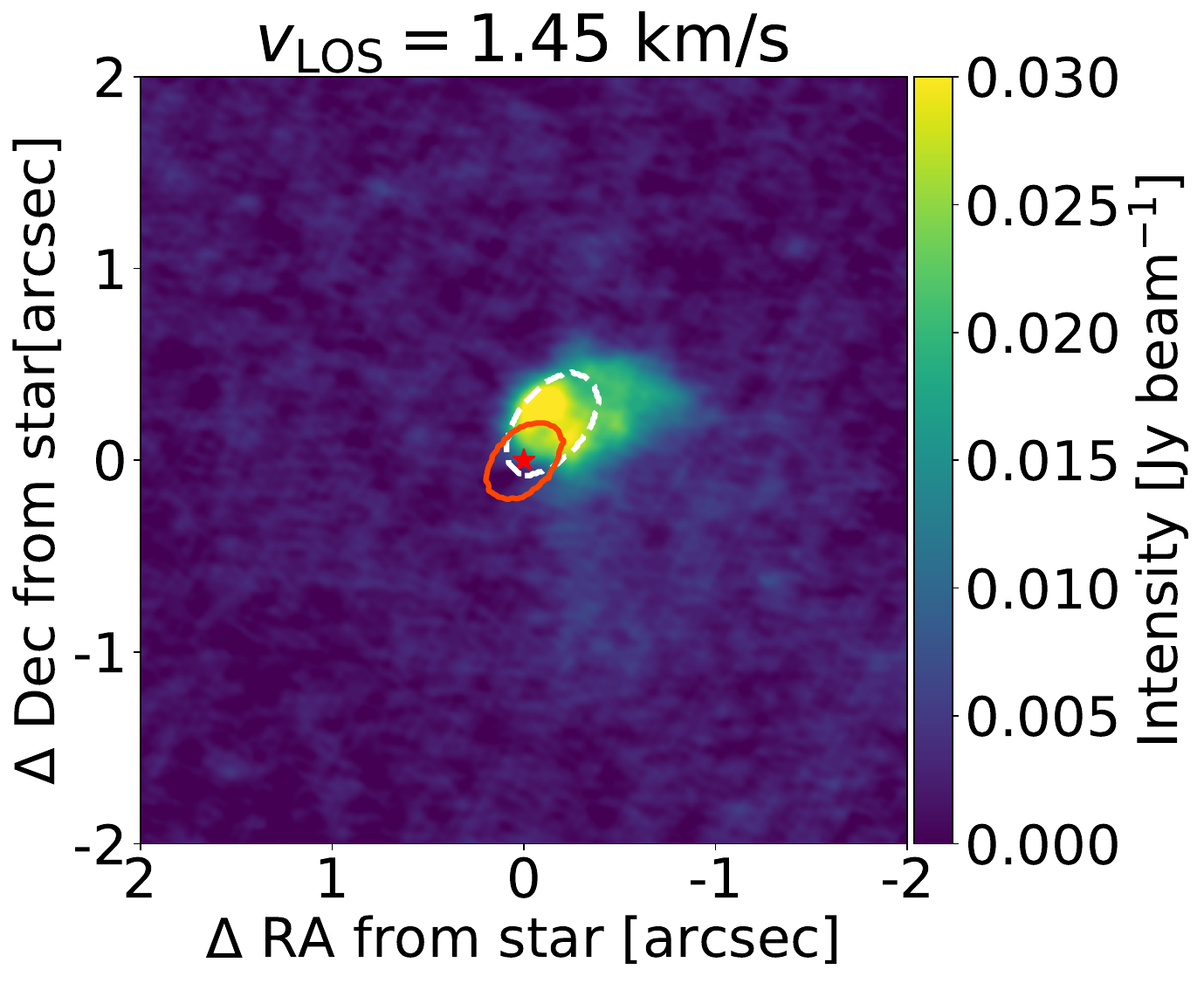}
\includegraphics[width=0.32\textwidth]{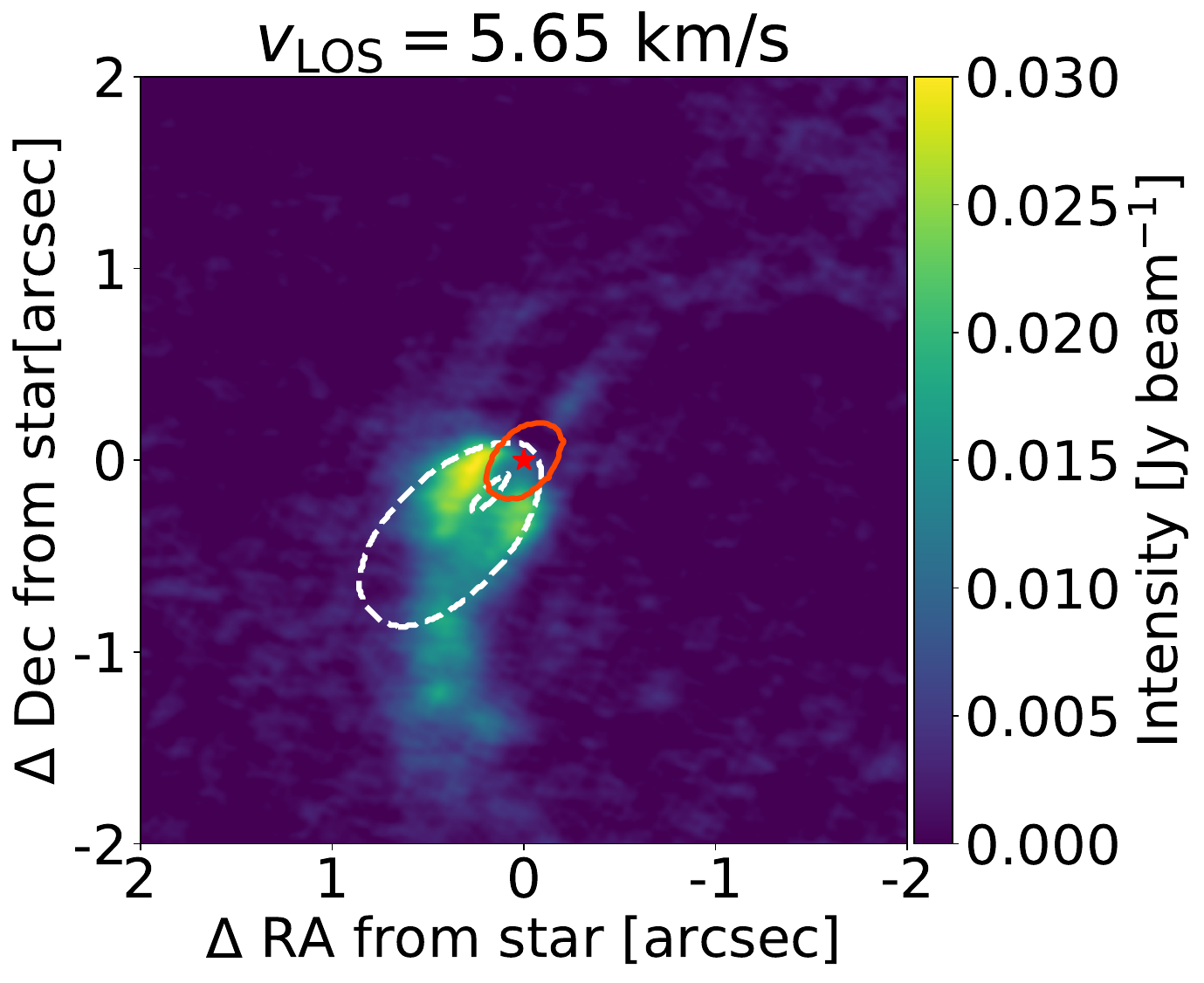}
\includegraphics[width=0.32\textwidth]{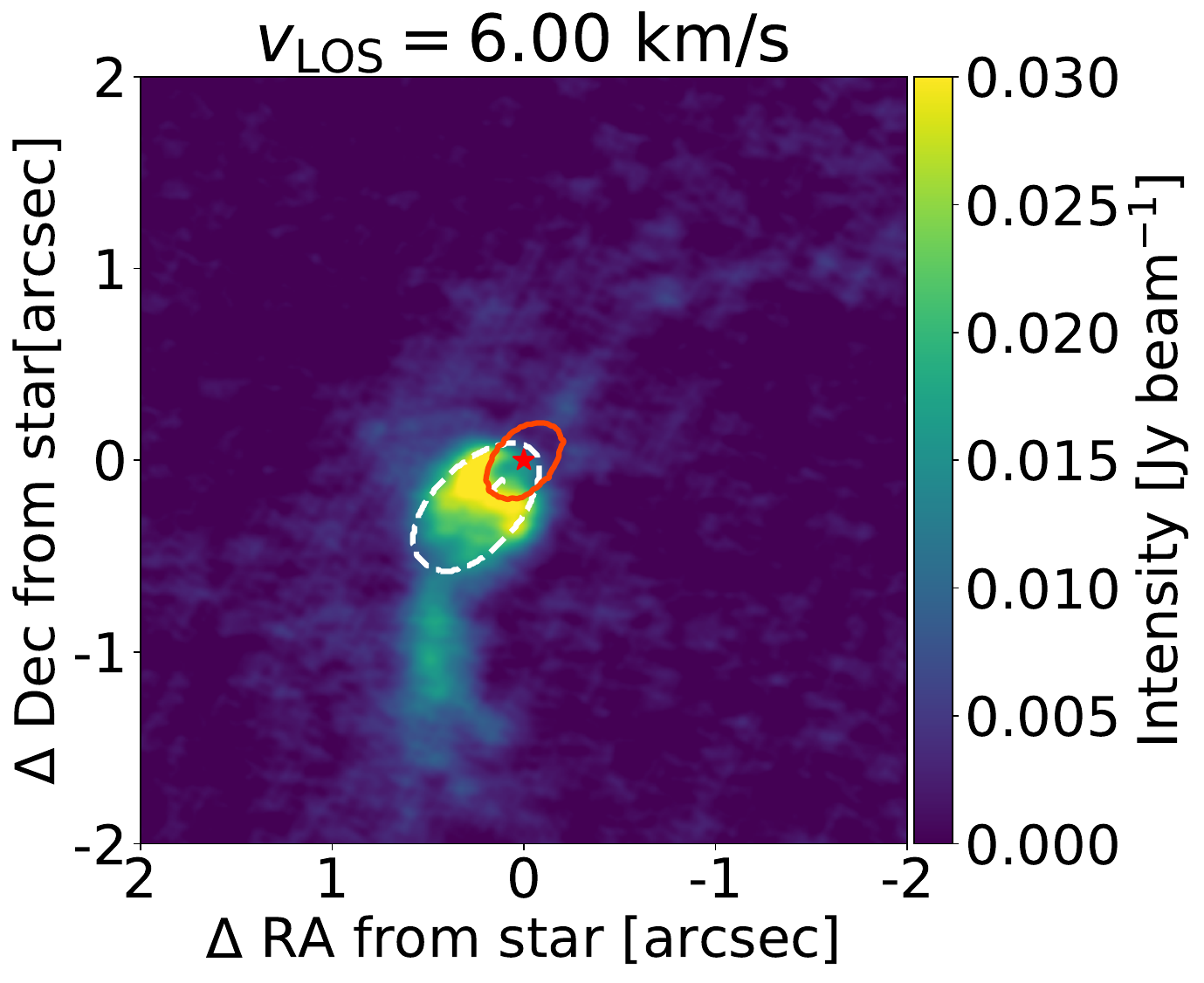}
\includegraphics[width=0.32\textwidth]{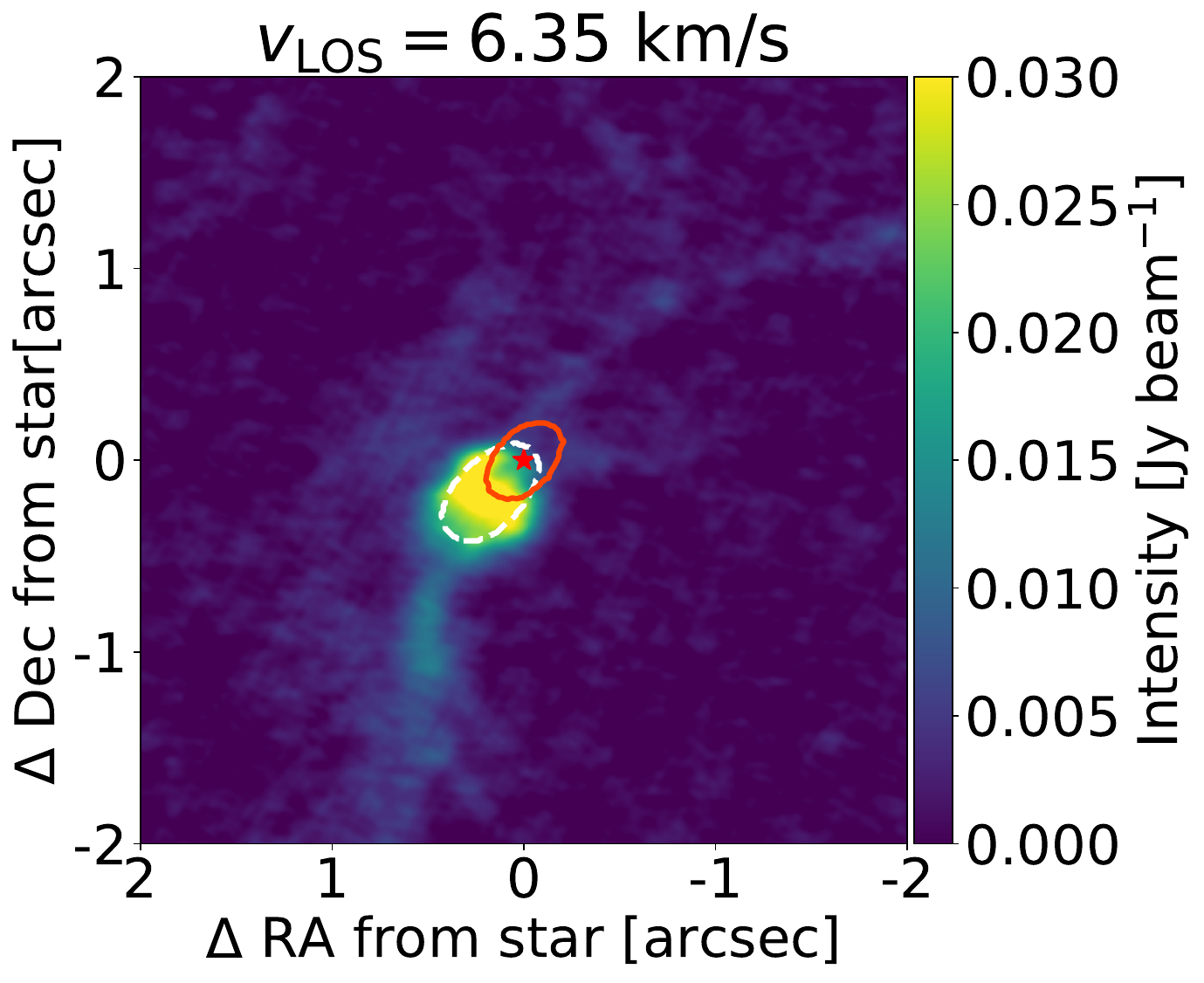}
\includegraphics[width=0.32\textwidth]{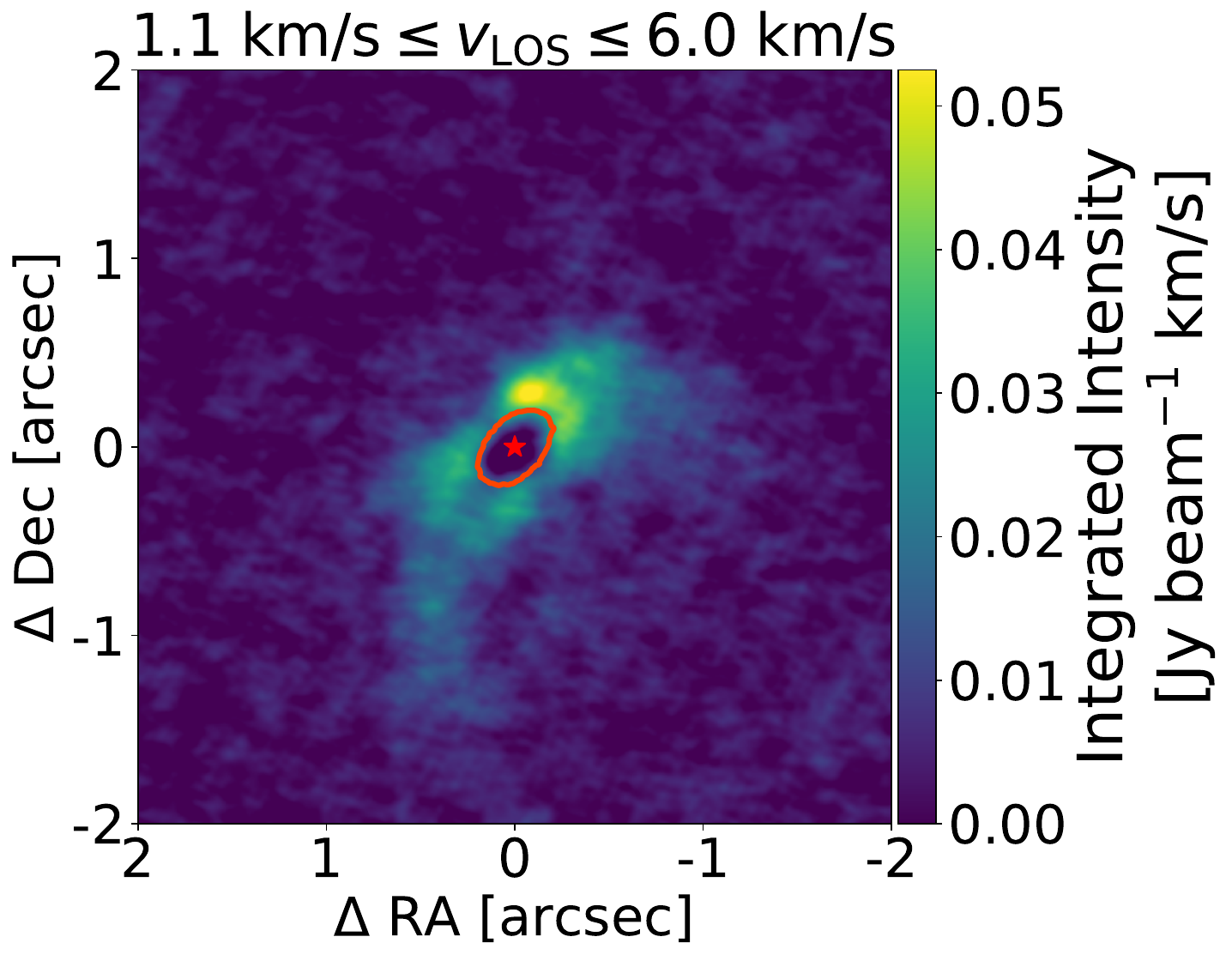}
\includegraphics[width=0.32\textwidth]{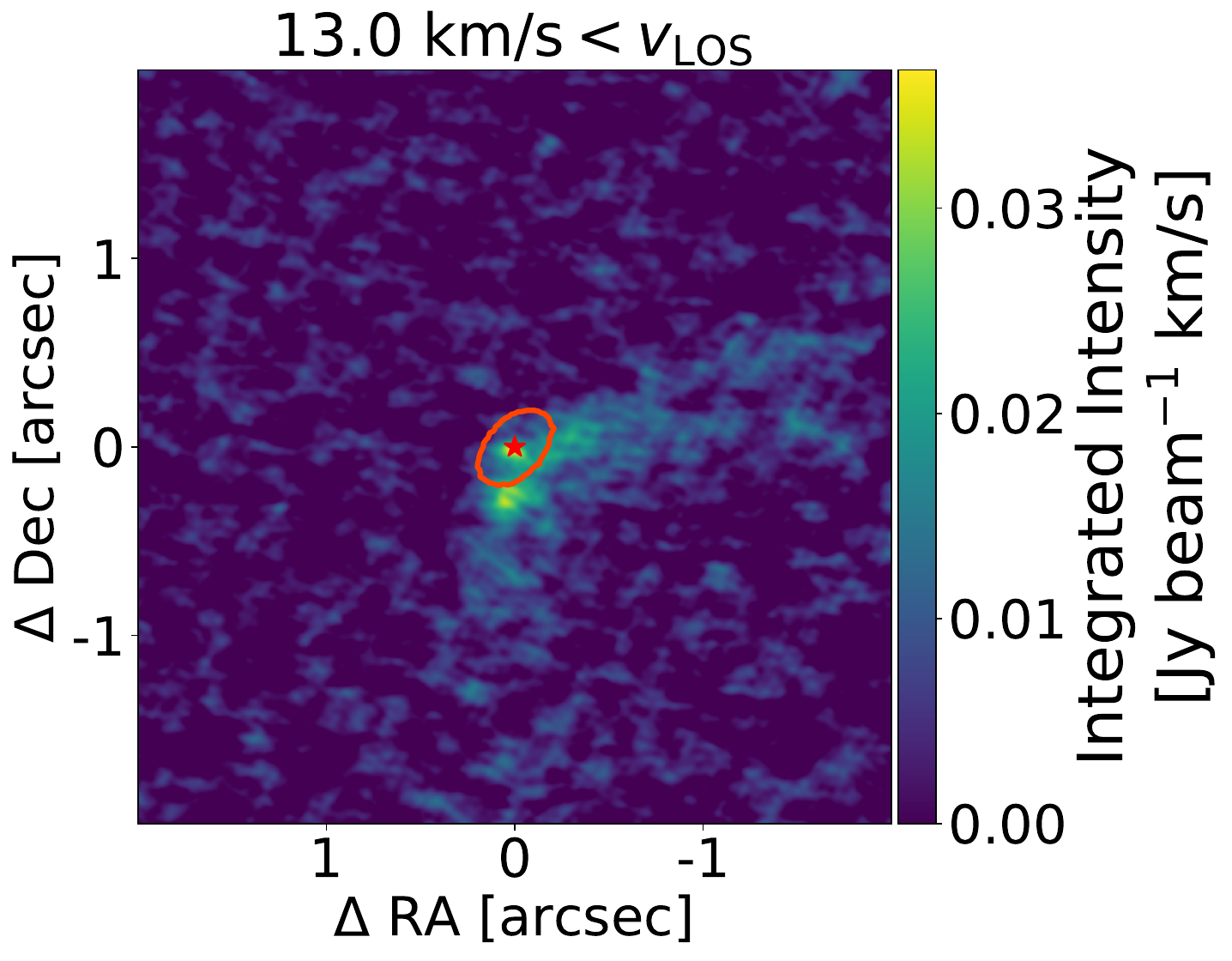}
\caption{ Close-up view of deformed protoplanetary disk around WSB 52. Zoomed-in views of selected channel maps are presented in the upper six panels. The range of the color bar is limited to 0-3 mJy/beam. White dotted line in the channel maps denotes the iso-velocity contours of the Keplerian disk with fiducial parameters. The lower two panels show the velocity-integrated map of the disk, with $1.1 \text{ km/s} \leq v_{\rm LOS} \leq 6.0 \text{ km/s}$ and $  13 \text{ km/s} < v_{\rm LOS} $. \aizw{Zoomed-in views of the continuum emission contour at a level of 0.1~mJy/beam are presented.}  In each image, the color scale ranges from 0 up to the maximum value in the respective image. }
\label{fig:channelszoom}
\end{figure*}

Beyond the shock boundary, the protoplanetary disk of WSB 52 is distinctly resolved, exhibiting clear spatial and kinematic separation from the shock front. Figure \ref{fig:channelszoom} provides channel maps with a zoomed-in view of the stellar disk. The disk appears to be deformed. The alignment of the \aizw{shock} front with the direction of disk deformation suggests that the shock front of the expanding bubble is colliding with and perturbing the protoplanetary disk. 

\aizw{On the other hand, the dusty disk—significantly more compact than the gas disk as shown in Figure \ref{fig:channelszoom}—appears unaffected by the shock. The disk is overall optically thick and exhibits only marginal substructure in the DSHARP sample \citep{huang2018}, with possible shallow gap/ring pairs at 13/17 and 21/25 au \citep{jennings2022}. Its extent corresponds to a high-density region where deformations are less likely, as observed in the inner regions of the gas disk. }

To quantify the deformation, we compare a Keplerian disk model \citep{2020zndo...4321137T,2021ApJS..257....2C} with the observed disk. In the implementation, we use our code for the Keplerian mask, which is publicly available at  \url{https://github.com/rorihara/Keplerian\_Mask\_Generator} and overall faster than an existing code, \texttt{keplerian\_mask} at  \url{https://github.com/richteague/keplerian\_mask} \citep{2020zndo...4321137T}.

We assume an axisymmetric disk. The height of the disk, $h(\rho)$, is specified by the following model: 
\begin{equation}
h(\rho) = h_0 \left( \frac{\rho}{100 \, \text{au}} \right)^p, 
\end{equation}
where $h_0$ determines the characteristic height for the model, $p$. Here, we only consider the upper layer of the disk, given that we only see that side in the observed disk. 

The velocity field of the disk model $v_{\rm model}$ is assumed to follow the Keplerian motion, and an additional systemic velocity  $v_{\text{sys},\star}$ is applied as an offset along the line-of-sight direction. Given each observational velocity $v_{\rm chan}$ and the channel width $\Delta v_{\rm chan}$, we make the masked region with $|v_{\rm model} - v_{\rm chan}| < \Delta v_{\rm chan}$. Here, we ignore the intrinsic line widths for the simplicity. 

In the modeling, we adopt the stellar mass and the disk parameters in Table \ref{table:parameters}. We assume the rotation direction of the disk to be consistent with the channel maps. The outer disk radius is assumed to be $150$ au. Due to significant cloud contamination in the relevant channels, the star's systemic velocity $v_{\text{sys},\star}$ remains highly uncertain. As the reference value we assume it to be $3.9$ km/s, but it may vary 0.5-1.0 km/s based on the observed maps. The assumed parameters are summarized in Table \ref{table:model_parameters}. 

The dashed lines in Figure \ref{fig:channelszoom} present a Keplerian disk model, which delineates the iso-velocity surfaces of the disk's flared upper layer as described. We present results using $h_0 = 5$ au and $p = 1$, demonstrating that the observed curvature of the disk cannot be accounted for by a simple flared disk model. We also test other parameters but consistently observe significant discrepancies, indicating intrinsic deformations likely caused by external forces.

Additionally, we identify high-velocity components near the disk reaching up to approximately $v_{\rm LOS} = 18$~km/s, some of which exceed the star's escape velocity of $\sim 3$~km/s at 100~au. This suggests ongoing mass loss from the disk. Figure \ref{fig:channelszoom} includes an integrated image highlighting regions with $13$~$\text{km/s}<v_{\rm LOS}$.

\section{Discussion} \label{sec:discuss}

\subsection{The proposed scenario for the observation} \label{sec:scenario}
As described in Section 3.1, the shell-like patterns can be well explained by the expanding bubble model\aizw{, which is distinct from sideways motions of bow shocks perpendicular to the jet directions \citep[e.g.,][]{Tafalla2017} and from outflow phenomena \citep[e.g,][]{harada2023}.} The similar bubbles have been observed in XZ Tau \citep{Krist1997,Krist1999,Kirst2008} and SVS 13 \citep{Hodapp2014}. Previous studies \citep{Kirst2008,Hodapp2014,gardner2016} propose that such bubbles result from the compression of cold gas, possibly ejected during prior outflow events, by powerful jet outbursts. The compressed hot gas exerts high pressure, leading to a spherical expansion that sweeps up interstellar material, manifesting as a shell. We postulate that the similar jet-driven mechanisms are responsible for the observed expanding bubble in WSB 52. 

\aizw{However, we note that similar CO bubble events have not been identified in other stellar systems, including XZ Tau and SVS 13. Although ALMA observations have revealed circumstellar dust and gas in these systems \citep{ichikawa2021,rodriguez2022,bianchi2023} and an outflow in XZ Tau \citep{zapata2015}, no corresponding CO emission from the bubbles has been detected. We therefore caution that the origin of the bubble in WSB 52 remains uncertain and warrants further investigation. The observed difference may be related to the fact that WSB 52 is a single star, while XZ Tau and SVS 13 are binary systems, but the effect of binarities is unclear. Nonetheless, this paper focuses on the outcome of the bubble's interaction with the disk rather than its formation mechanism, and its conclusions do not depend on any specific assumption regarding the bubble's origin.}

\aizw{The} key distinction between the bubbles observed in XZ Tau/SVS 13 and the current bubble in WSB 52 lies in their velocity profiles. The radial expansion velocity of the WSB 52 bubble ($12.5$~km/s) exceeds its velocity relative to the star, enabling a returning motion toward the star. In contrast, previously reported bubbles in XZ Tau and SVS 13 primarily exhibit outward discharge without significant returning motion. This returning movement in WSB 52 facilitates interaction between the bubble and the circumstellar material. The observed brightness distribution of the bubble sphere in Figure \ref{fig:channels_maps_all} appears asymmetric: the side close to the star is bright, while the side far from the star remains dark. This asymmetry is consistent with a greater amount of circumstellar material near the star, where strong interactions with the bubble have occurred.

We also explore whether the jets could supply the energy driving the current bubble. The jet mass ejection rates for Class II T Tauri stars have been found to range from $10^{-9}$ to $10^{-7}$~$M_{\odot}/{\rm yr}$ \citep{Ellerbroek2013}.  Furthermore, several systems have been reported to exhibit episodic ejections on timescales of around 10 years and rates exceeding $10^{-8}$~$M_{\odot}/{\rm yr}$ \citep{takami2023,pyo2024}. A jet outburst characterized by a radial velocity of 200~km/s and a mass ejection rate of $10^{-7}$~$M_{\odot}/{\rm yr}$ yields a kinetic energy rate of approximately $0.4 \times 10^{41}$~erg/yr. This energy output is sufficient to drive an expanding bubble with kinetic energy in the range of $0.4-4 \times 10^{41}$~erg within one to ten years. In the current case, the mass accretion rate for WSB 52 was estimated to be $10^{-7.6 \pm 0.5}, M_{\odot}/\mathrm{yr}$ \citep{Natta2006,andrews2018}, implying that a jet mass ejection rate can be as high as $10^{-8}-10^{-7}$~$M_{\odot}/\mathrm{yr}$ \citep{Ellerbroek2013}. Given the temporal variability of jet activity, it is reasonably that jets were more powerful in the past, thereby driving the bubble formation. 

\aizw{Figure \ref{fig:summary}} summarizes the proposed scenario accounting for these features from the formation of the bubble to the current state. We propose that a uniform explosion is sweeping through the circumstellar material, thereby perturbing and shaking or even destructing the disk structure.

\subsection{Simple model for disk deformation by bubble expansion}
As observed, the disk is deformed along the expansion direction of the bubble. This deformation may be explained by considering the ram pressure exerted by the bubble on the disk, in a manner similar to the interaction between supernova ejecta and a protoplanetary disk \citep[e.g.,][]{Chevalier2000,Ouellette2007}. Assuming a gas flow with density \(\rho_{\rm flow}\) and the velocity $v_{\rm flow}$, the ram pressure \(\rho_{\rm flow} v_{\rm flow}^2\) acts on the disk, which has a surface density \(\Sigma_{\rm disk}\). The corresponding equation of motion is
\begin{equation}
    \rho_{\rm flow} v_{\rm flow}^2 = \Sigma_{\rm disk} a, 
\end{equation}
where $a$ is the acceleration.
Assuming that the acceleration is maintained for a duration \(\Delta t\), the surface density of the impacting flow is given by
\begin{equation}
    \Sigma_{\rm flow} = \rho_{\rm flow} v_{\rm flow} \Delta t.
\end{equation}
Thus, the vertical velocity \(v_{\zeta}\) acquired by the disk is
\begin{equation}
    v_{\zeta} = a \Delta t = \frac{\Sigma_{\rm flow}}{\Sigma_{\rm disk}} v_{\rm flow},
\end{equation}
and the resulting vertical displacement \(D_{\zeta}\) is approximated by
\begin{equation}
    D_{\zeta} = \frac{1}{2} a \Delta t^2 = \frac{\Sigma_{\rm flow}}{2 \Sigma_{\rm disk}}\ v_{\rm flow} \Delta t.
\end{equation}

The surface density associated with the bubble is expressed as
\begin{equation}
    \Sigma_{\rm flow} = \frac{M_{\rm bubble}}{4 \pi r_{\rm bubble}^2},
\end{equation}
where \(r_{\rm bubble}\) is the bubble radius and \(M_{\rm bubble}\) is its mass.

\begin{figure*}[t]
\centering
\includegraphics[width=0.60\textwidth]{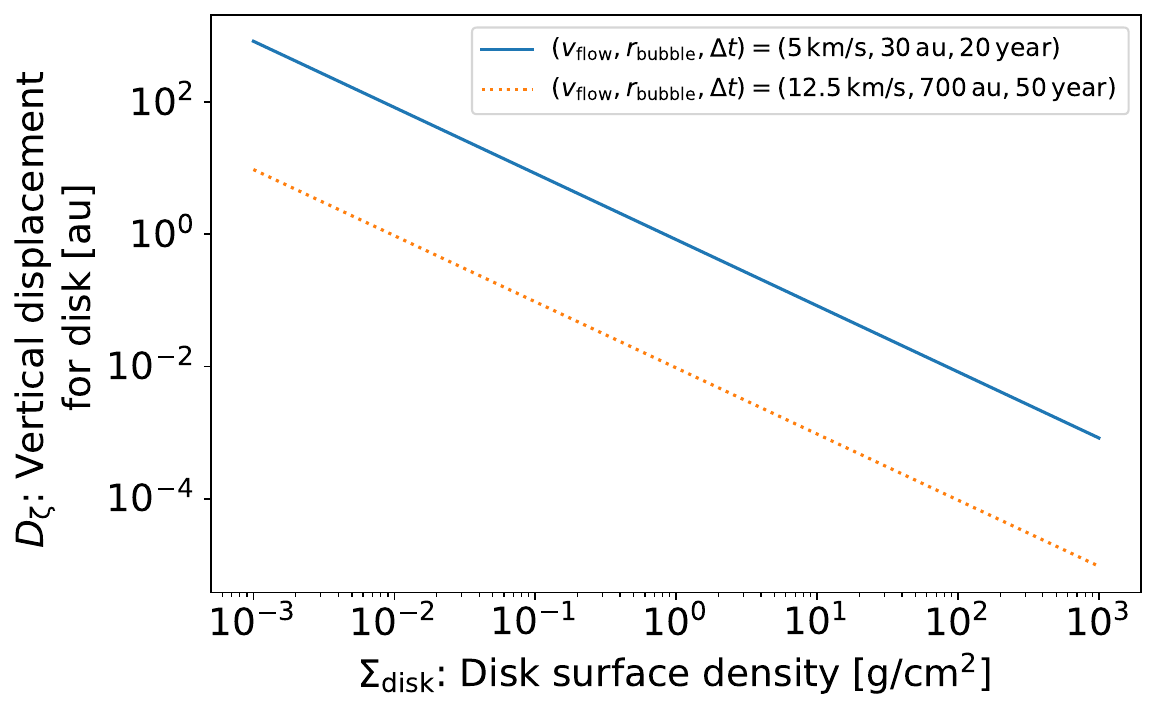}
\caption{Vertical displacement for disk caused by bubble pressure versus disk surface density. The solid line assumes $(v_{\rm flow}, r_{\rm bubble}, \Delta t) = (5 \,\text{km/s}, 30 \,\text{au}, 20\, \text{yrs})$, while the dotted line assumes $(v_{\rm flow}, r_{\rm bubble}, \Delta t) = (12.5\, \text{km/s}, 700 \,\text{au}, 50 \,\text{yr})$. }
\label{fig:disp_density}
\end{figure*}

Consequently, the vertical displacement becomes
\begin{align}
    D_{\zeta} &= 0.01\,\text{au} \left(\frac{M_{\rm bubble}}{10^{-4}\,M_{\odot}}\right) \left(\frac{r_{\rm bubble}}{700\,\text{au}}\right)^{-2} \nonumber\\
    &\quad \times \left(\frac{\Sigma_{\rm disk}}{1\,\text{g/cm}^2}\right)^{-1}
    \left(\frac{v_{\rm flow}}{12.5\,\text{km/s}}\right)
    \left(\frac{\Delta t}{50\,\text{yr}}\right).
\end{align}
Figure~\ref{fig:disp_density} illustrates the result with $(v_{\rm flow}, r_{\rm bubble}, \Delta t) = (12.5 \,\text{km/s}, 700\,\text{au}, 50\,\text{yr})$, where $v_{\rm flow}=12.5 \text{km/s}$ represents the bubble expansion velocity, $r_{\rm bubble}=700\,\text{au}$ is the current bubble radius, and $\Delta t$ is taken to be 50 yr, approximately equal to the shell's crossing time over 1 arcsec. Under the assumption \(\Sigma_{\rm disk}=1\,\text{g/cm}^2\), the vertical displacement is approximately 0.01 au. This value is likely negligible and inconsistent with the observed significant deformation.

If, however, the bubble-disk interaction began at an earlier phase when the disk and bubble centers were closer, we can expect a larger vertical displacement. In that case, the bubble's relative velocity with respect to the disk center should be considered, and the bubble radius is not constant. Assuming a relative velocity of $v_{\rm rel}= 7.5$ km/s ($\simeq$ 1.5 au/yr), as estimated from the current separation between the two centers, the effective flow velocity is $v_{\rm flow} = u_{\rm bubble} - v_{\rm rel} = 5$ \text{km/s}. Ignoring the time variability of the bubble radius, the vertical displacement can be then  estimated as: 
\begin{align}
    D_{\zeta} &= 0.8 \,\text{au} \left(\frac{M_{\rm bubble}}{10^{-4}\,M_{\odot}}\right) \left(\frac{r_{\rm bubble}}{30\,\text{au}}\right)^{-2} \nonumber\\
    &\quad \times \left(\frac{\Sigma_{\rm disk}}{1\,\text{g/cm}^2}\right)^{-1}
    \left(\frac{v_{\rm flow}}{5\,\text{km/s}}\right)
    \left(\frac{\Delta t}{20\,\text{yr}}\right).
\end{align}
Here, we adopt fiducial values with  $(v_{\rm flow}, r_{\rm bubble}, \Delta t) = (5\,\text{km/s}, 30\,\text{au}, 20\,\text{yr} \,(\simeq 30 \, \text{au}/v_{\rm rel}))$, and the expected displacement is shown in Figure~\ref{fig:disp_density}. Assuming a lower surface density, a smaller effective radius, or increased bubble mass (or energy), the resulting displacement can become significantly larger.
Overall, the feedback on the disk is expected to be weaker in the dense inner regions and stronger in the less-dense outer regions, which is consistent with the observation that the deformation increases with radius\aizw{, while no significant deformation is observed in the compact dusty disk.}

Although the current model provides a preliminary interpretation of the disk deformation, further detailed modeling of the observed vertical displacement can offer valuable constraints on the evolution of the bubble and the surface density profile of the disk.

\subsection{Implications for star and planet formation}

Jets and outflows play a critical role in removing excess angular momentum from accreting materials by expelling mass flux \citep{Blandford1982}. Their substantial kinetic energy is injected into surrounding molecular clouds, significantly influencing the interstellar environment. Herbig-Haro objects, luminous nebulae at jet shock fronts, serve as direct evidence of such energy injections \citep[e.g.,][]{Reipurth2001}. This energy input at larger scales increases turbulence and pressure within surrounding environments, and thereby regulating star formation efficiency  \citep[e.g.,][]{Norman1980,Nakamura2007}. Our findings extend this impact to the protoplanetary disk scale, demonstrating that jets can interact with circumstellar environments through expanding bubbles. The observed disk deformation indicates that the effects of such jet-driven bubbles are substantial, particularly in the outer disk regions. 

Beyond accretion from larger structures like envelopes \aizw{\citep[e.g.,][]{ohashi2014,sakai2014,yen2014}}, several mechanisms transfer energy and materials into protoplanetary disks. Proplyds, or ionized disks, are heated by radiation from nearby massive stars, as seen in the Orion Nebula Cluster \citep[e.g.,][]{odell1993,bally2000}. Supernovae can reshape the structure of protoplanetary disks while injecting short-lived radioactive isotopes, such as $^{26}$Al and $^{60}$Fe, which influence the thermal evolution of forming planetary systems \citep[e.g.,][]{Chevalier2000,Ouellette2007}. Outbursts from central stars, such as FU Ori-type events, can also reshape disk structures by altering the temperature distribution, which shifts snowline positions \citep{Banzatti2015,cieza2016,lee2019}. Our study reveals a novel mechanism of energizing a disk through jet-induced bubbles, especially influencing the vertical structure in localized environments.

The deformation of the disk shape provides clear evidence of significant kinetic energy injection into the system. This interaction suggests that part of the disk mass could be lost due to the strong momentum transfer from the bubble. The vertical extent and dynamics of the gas may be enhanced by the explosive event, altering turbulence levels through shock interactions. Shock heating could also energize dust grains, potentially leading to \aizw{molecular desorption}, which could produce observable signatures\aizw{—for example, through the emergence of specific shock tracers.}

\section{Conclusion} \label{sec:conclusion}
 
In this study, we report the first case for the jet feedback on the disk via an expanding bubble. Specifically, jets generate bubbles, which interact with disks. The disks again feed the jets. We name this process as ``\textit{jet-bubble-disk interaction}''. 

The current analysis solely relies on \aizw{$^{12}\mathrm{CO}\,(J=2\mbox{--}1)$}. Therefore, follow-up observations of CO isotopes are highly desirable to gain a better understanding of the mass and temperature distribution. Additionally, follow-up observations for other chemical species are helpful for investigating the chemical compositions of the gas, which may have been altered by the bubble interaction. Furthermore, detailed dynamical modeling of the disk's deformation caused by bubble wind will be important for understanding the bubble's effect on the disk.

In our analysis, among the DSHARP targets, only WSB 52 exhibits similar events, suggesting that such explosive interactions may be rare. \aizw{The estimation of occurrence rates of bubbles is challenging due to the uncertain frequency of jet-interacting cold material. Such cold material may be ejecta from previous outflows or small-scale cloudlets that are inherent in parent molecular clouds \citep[e.g.,][]{koyama2000,hennebelle2012}.} To better understand the prevalence and impact of these bubbles, more extensive surveys for similar explosive events are highly recommended. For example, systems with higher mass accretion rates—such as protostars and FU Ori-type stars—could exhibit more energetic phenomena. In addition, theoretical modelings are necessary to better understand the mechanisms underlying jet-induced bubble formation and their long-term effects on protoplanetary disks. 


\section*{Acknowledgements}
\aizw{We thank an anonymous referee for constructive comments.} This work was supported by JSPS KAKENHI grant number 22H01274. Data analysis was in part carried out on the Multi-wavelength Data Analysis System operated by the Astronomy Data Center (ADC), National Astronomical Observatory of Japan. This paper makes use of the following ALMA data: ADS/JAO.ALMA \#2015.1.00486.S, ADS/JAO.ALMA \#2015.1.00964.S. ALMA is a partnership of ESO (representing its member states), NSF (USA) and NINS (Japan), together with NRC (Canada), MOST and ASIAA (Taiwan), and KASI (Republic of Korea), in cooperation with the Republic of Chile. The Joint ALMA Observatory is operated by ESO, AUI/NRAO and NAOJ.  This work has made use of data from the European Space Agency (ESA) mission
{\it Gaia} (\url{https://www.cosmos.esa.int/gaia}), processed by the {\it Gaia}
Data Processing and Analysis Consortium (DPAC,
\url{https://www.cosmos.esa.int/web/gaia/dpac/consortium}). Funding for the DPAC
has been provided by national institutions, in particular the institutions
participating in the {\it Gaia} Multilateral Agreement.

\section*{Availability of data and materials} 
\aizw{$^{12}\mathrm{CO}\,(J=2\mbox{--}1)$ image cube is provided in \url{https://doi.org/10.5281/zenodo.15314468}.} 
The raw ALMA data are publicly available via the ALMA archive https://almascience.nrao.edu/aq/. The DSHARP data used in this article are available in the DSHARP Data Release at https://bulk.cv.nrao.edu/almadata/lp/DSHARP. 

\section*{Code availability} The original code used for reconstrucing the images is available in the DSHARP Data Release at https://bulk.cv.nrao.edu/almadata/lp/DSHARP. The Keplerian mask code is publicly available at \url{https://github.com/rorihara/Keplerian\_Mask\_Generator}. 

{\it Software}:    {\tt Astropy} \citep{astropy2013,2018AJ....156..123A,2022ApJ...935..167A}, {\tt CASA} \citep{mcmullin2007}, {\tt CARTA} \citep{comrie2021carta}, {\tt Jupyter Notebook} \citep{kluyver2016}, {\tt Matplotlib} \citep{hunter2007}, {\tt NumPy} \citep{walt2011}, {\tt Pandas} \citep{mckinney-proc-scipy-2010}, {\tt SciPy}  \citep{virtanen2020}. 

\appendix

\section{Estimation of mass and kinematic energy of bubble using $^{12}$CO emission line \label{sec:mass_energy_bubble}} 
We perform an order-of-magnitude estimation of the bubble's mass and kinematic energy to assess whether the observed disk can be plausibly energized by the stellar jet. For this estimation, we utilize the intensities of $^{12}$CO emission line, assuming that the line emission is not fully optically thick.  

Assuming the local thermal equilibrium (LTE), we can derive the molecular column density from the observed optical depth as follows \citep{mangum2015,orihara2023}:  
\begin{eqnarray}
    N_{\rm{}tot}=&\cfrac{3h}{8\pi^3|\mu_{\rm{}lu}|^2}\cfrac{Q_{\rm{}rot}}{g_{\rm{}u}}\exp\Big(\cfrac{E_{\rm{}u}}{kT_{\rm{}ex}}\Big) 
    \Big[\exp\Big(\cfrac{h\nu}{kT_{\rm{}ex}}\Big)-1\Big]^{-1}\int\tau_{\rm{}g}dv,
\end{eqnarray}
where \( N_{\rm{}tot} \) is the total column density, \( h \) is the Planck constant,  \( E_{\rm{}u} \) is the energy of the upper energy level, \( k \) is the Boltzmann constant, \( T_{\rm{}ex} \) is the excitation temperature, which is assumed to be equal to the gas temperature in the LTE analysis, \( \nu \) is the frequency of the transition, and \( \int \tau_{\rm{}g}dv \) is the integrated optical depth. 

The square of the transition dipole moment, \( |\mu_{\rm{}lu}|^2 \), is given by:
\begin{eqnarray}
    |\mu_{\rm{}lu}|^2=\mu^2 \frac{J_{\rm{}u}^2}{J_{\rm{}u}(2J_{\rm{}u}+1)},
\end{eqnarray}
where \( \mu \) $(= 0.110 \times 10^{-18}$~esu cm for $^{12}$CO) is the permanent dipole moment, and  \( J_{\rm{}u} \) is the rotational quantum number of the upper energy level. The degeneracy of the upper energy level, \( g_{\rm{}u} \), is given by:
\begin{eqnarray}
    g_{\rm{}u}=2J_{\rm{}u}+1,
\end{eqnarray}
and the energy of the upper energy level, \( E_{\rm{}u} \), is given by:
\begin{eqnarray}
    E_{\rm{}u}=hBJ_{\rm{}u}(J_{\rm{}u}+1),
\end{eqnarray}
where \( B \) is the rotational constant $(= 57635.96$ MHz for $^{12}$CO). The partition function \( Q_{\rm{}tot} \) is given by 
\begin{eqnarray}
  Q_{\rm{}rot} = \sum_{i} g_{i} \exp\left(-\frac{E_{i}}{kT}\right), 
\end{eqnarray}
where the value is approximated by summing up to $J_{i}<20$ in this study. The optical depth is derived from the observed intensity $I_\nu = I_{\rm obs}$ and the 
Planck function 
$B_\nu (T_{\rm g})$ using the following relation: 
\begin{equation}
I_\nu = B_\nu (T_{\rm g})  (1 - e^{-\tau_{\rm g}}). 
\end{equation}
Specifically, we use
\begin{equation}
\tau_{\rm g} = - \ln \left(1 - \frac{ I_\nu}{ B_\nu (T_{\rm g}) } \right).
\end{equation}
With the derived column density $N_{\rm{tot}}$, the gas column density is then calculated by
\begin{eqnarray}
    \Sigma_{\rm{}g} =\frac{m_{\rm{H_2}}N_{\rm{tot}}}{X}, \label{eq:sigma_mass}
\end{eqnarray}
where $m_{\rm{H_2}}$($=3.32\times10^{-24}~\rm{g}$) is the molecular mass of $\rm{H_2}$, and $X$ is the molecular abundance ratio with $\rm{}H_2$. We assume $X=10^{-4}$ for $^{12}$CO. 

The mass is derived by integrating Eq. (\ref{eq:sigma_mass}) over each pixel, using the estimates of $\tau_{\rm g}$ obtained from the real cube data. The integration region is defined between two circles corresponding to the bubble models with radii $r_{\rm bubble} = 4$ and $7.0$ arcsec. To remove the contribution from disk emission, we exclude intensities within 1.5 arcsec from the star. We also remove channels with $v_{\rm LOS}$ between 1.45 and 6 km/s to eliminate cloud contamination. The mass depends on the unknown gas temperature, so we vary the temperature from $25$~K to $95$~K in steps of $10$~K.

During the analysis, we notice that $I_\nu$ and $\tau_{\rm g}$ can take negative values because of random noises, thus possibly biasing the estimation of $\tau_{\rm g}$. To understand the possible uncertainty in the mass, we employ three different integration methods : (1) Integrating by forcing $I$ to be zero if it is negative, (2) Integrating only in regions where $I$ is above the 3-sigma level, and (3) Integrating without applying any specific conditions, accounting for both positive and negative fluxes. To reduce the biases, we also smooth the cube data using a beam size of 0.25 arcsec reducing the number of negative pixels.  As a result, we find that the mass range is $0.2 - 1.1 \times 10^{-4} M_{\odot}$.

The kinematic energy is then estimated from the total mass and the velocity as follows: 
\begin{equation}
E_{\rm shell} = \frac{1}{2} m_{\rm g} u_{\rm bubble}^{2}. 
\end{equation}
Assuming $u_{\rm bubble} = 12.5$~km/s and the mass range, we find $E_{\rm kin}$ varies between $0.3-1.6 \times10^{41} $~erg. The method (1) gives $E_{\rm kin}=0.8-1.6 \times10^{41}$~erg, and the other two give the lower total energy, $E_{\rm kin}=0.3-1.0 \times10^{41}$~erg. As discussed in the main text, this energy can be explained by the energy injection though the stellar jet.


\section{Analytical solution for iso-velocity section for shock boundary model} \label{sec:analytical_shock}

For the analysis in Sec \ref{sec:shock_boundary}, we derive an analtycial expression for the model's projection onto the observational $(x,y)$ coordinate system with fixed $v_{\rm LOS}$. We denote the unit vector from the bubble center to the star by $\mathbf{\hat{\zeta}} = (\hat{\zeta}_{x}, \hat{\zeta}_{y}, \hat{\zeta}_{z})^{T}$. To make the coordinate system, we prepare two orthogonal unit vectors as follows:
\begin{eqnarray}
\mathbf{\hat{\xi}} &\equiv& (\hat{\xi}_{x}, \hat{\xi}_{y}, \hat{\xi}_{z})^{T} = \frac{1} { \sqrt{ \hat{\zeta}_x^{2} +\hat{\zeta}_y^{2}}} (-\hat{\zeta}_{y}, \hat{\zeta}_{x}, 0)^{T}, \\
\mathbf{ \hat{\eta}} &\equiv& (\hat{\eta}_{x}, \hat{\eta}_{y}, \hat{\eta}_{z})^{T} = \mathbf{ \hat{\zeta}} \times \mathbf{\hat{\xi}} = \frac{1} { \sqrt{ \hat{\zeta}_x^{2} +\hat{\zeta}_y^{2}}} (-\hat{\zeta}_{z} \hat{\zeta}_{x}, -\hat{\zeta}_{z} \hat{\zeta}_{y}, \hat{\zeta}_{x}^{2} + \hat{\zeta}_{y}^{2}) ^{T}, 
\end{eqnarray}
where $\mathbf{\hat{\xi}}$ and $\mathbf{\hat{\eta}}$ align with $\xi$ and $\eta$ axes, respectively.

The position on the shock boundary $\mathbf{S}$ measured from the bubble center is given by 
\begin{equation}
\mathbf{S} = (d + \zeta_{\rm surf}(\rho))\mathbf{\hat{\zeta}} + (\rho \cos \phi)\mathbf{\hat{\xi}} + (\rho \sin \phi) \mathbf{ \hat{\eta}}, \label{eq:boundary_pos}
\end{equation}
where $\phi$ is the azimuthal angle for $(\xi, \eta)$, ranging from $0$ to $2\pi$: 
\begin{align}
    \xi &= \rho \cos \phi,  \\
    \eta &= \rho \sin \phi.  
\end{align}
$\rho$ can vary within the range $[0, \rho_{\rm inter}]$. 

The line-of-sight velocity $v_{\rm LOS}$  for the boundary model is  given by 
\begin{align}
v_{\rm LOS} &= u_{\rm bubble}\frac{\mathbf{S}\cdot \mathbf{\hat{e}}_{z} }{||\mathbf{S}||} + v_{\rm sys, bubble}  \\
&= \frac{u_{\rm bubble}}{\sqrt{(d + \zeta_{\rm surf}(\rho))^{2} + \rho^{2}}}  \left[(d + \zeta_{\rm surf}(\rho))\hat{\zeta}_{z} + \left( \rho \sin \phi  \sqrt{ \hat{\zeta}_x^{2} +\hat{\zeta}_y^{2}}\right) \right] + v_{\rm sys, bubble}, 
\end{align}
where $\mathbf{\hat{e}}_{z}$ is the unit vector in the direction of the $z$-axis (the line-of-sight direction). We can solve $\phi$ from $v_{\rm LOS}$ and $\rho$ as follows: 
\begin{equation}
\phi(v_{\rm LOS}, \rho) = \arcsin \left [ \frac{1}{\rho \sqrt{\hat{\zeta}_{\rm x}^{2} + \hat{\zeta}_{\rm y}^{2}}} \left[ \frac{(v_{\rm LOS}- v_{\rm sys, bubble}) \sqrt{(d + \zeta_{\rm surf}(\rho))^{2} + \rho^{2}}}{u_{\rm bubble}} - (d + \zeta_{\rm surf}(\rho)) \hat{\zeta}_{z} \right]\right]. \label{eq:phi_los_rho}
\end{equation}
According to Eq (\ref{eq:boundary_pos}), the positions on the boundary $(x_{\rm surface}, y_{\rm surface})$ are given by
\begin{align}
x_{\rm surface} &= \mathbf{S} \cdot \mathbf{\hat{e}}_{x} 
=  (d + \zeta_{\rm surf}(\rho)) \hat{\zeta}_x + (\rho \cos \phi)\hat{\xi}_x + (\rho \sin \phi) \hat{\eta}_x,  \\
y_{\rm surface} &= \mathbf{S} \cdot \mathbf{\hat{e}}_{y} 
=(d + \zeta_{\rm surf}(\rho)) \hat{\zeta}_y + (\rho \cos \phi)\hat{\xi}_y + (\rho \sin \phi) \hat{\eta}_y.  \label{eq:y_position}
\end{align}
Using the above equations and Eqs (\ref{eq:phi_los_rho}-\ref{eq:y_position}), the iso-velocity contour for the boundary model at the observed velocity $v_{\rm LOS}$ can be obtained as follows: 
\begin{eqnarray}
x_{\rm contour}(v_{\rm LOS}, \rho)  &=& (d + \zeta_{\rm surf}(\rho))\hat{\zeta}_{x} + \frac{\left (- \rho \hat{\zeta}_{y} \cos \phi(v_{\rm LOS}, \rho) - \rho \hat{\zeta}_{z} \hat{\zeta}_{x}  \sin \phi(v_{\rm LOS}, \rho) \right)} { \sqrt{ \hat{\zeta}_x^{2} +\hat{\zeta}_y^{2}}} ,  \\
y_{\rm contour}( v_{\rm LOS}, \rho)  &=& (d + \zeta_{\rm surf}(\rho))\hat{\zeta}_{y}  + \frac{ \left ( \rho  \hat{\zeta}_{x} \cos \phi(v_{\rm LOS}, \rho) - \rho  \hat{\zeta}_{z} \hat{\zeta}_{y} \sin \phi(v_{\rm LOS}, \rho) \right)} { \sqrt{ \hat{\zeta}_x^{2} +\hat{\zeta}_y^{2}}}, 
\end{eqnarray}
where $\rho$ can vary within the range $[0, \rho_{\rm inter}]$.

\bibliographystyle{aasjournal}
\bibliography{ref}
\end{CJK*}
\end{document}